\documentclass{aa}  

\usepackage{graphicx}
\usepackage{txfonts}

\usepackage{amsmath}	
\usepackage{amssymb}	
\usepackage{mathtools}

\usepackage{color}
\usepackage{natbib,twoopt}
\usepackage[hyphenbreaks]{breakurl}
\usepackage[breaklinks]{hyperref}      
\bibpunct{(}{)}{;}{a}{}{,}             
\definecolor{cobalt}{rgb}{0.06, 0.2, 0.65}
\hypersetup{
  colorlinks,
  citecolor=blue,
  linkcolor=blue,
  urlcolor=blue,
}
\makeatletter
  \newcommandtwoopt{\citeads}[3][][]{\href{http://adsabs.harvard.edu/abs/#3}%
    {\def\hyper@linkstart##1##2{}%
     \let\hyper@linkend\@empty\citealp[#1][#2]{#3}}}
  \newcommandtwoopt{\citepads}[3][][]{\href{http://adsabs.harvard.edu/abs/#3}%
    {\def\hyper@linkstart##1##2{}%
     \let\hyper@linkend\@empty\citep[#1][#2]{#3}}}
  \newcommandtwoopt{\citetads}[3][][]{\href{http://adsabs.harvard.edu/abs/#3}%
    {\def\hyper@linkstart##1##2{}%
     \let\hyper@linkend\@empty\citet[#1][#2]{#3}}}
  \newcommandtwoopt{\citeyearads}[3][][]%
    {\href{http://adsabs.harvard.edu/abs/#3}
    {\def\hyper@linkstart##1##2{}%
     \let\hyper@linkend\@empty\citeyear[#1][#2]{#3}}}
\makeatother

\newcommand{\Msun}{M$_{\odot}$}
\newcommand{\Rsun}{R$_{\odot}$}

\definecolor{smalt(darkpowderblue)}{rgb}{0.0, 0.2, 0.6}
\definecolor{forestgreen(traditional)}{rgb}{0.0, 0.5, 0.0}

\defcitealias{RVJ}{RVJ}

\defcitealias{MD17}{MD17}

\newcommand{\am}{AM\,CVn binary}
\newcommand{\ams}{AM\,CVn binaries}

\newcommand{\cv}{CV}
\newcommand{\cvs}{CVs}

\newcommand{\hecv}{He\,CV}
\newcommand{\hecvs}{He\,CVs}

\newcommand{\ucxbs}{ultra-compact X-ray binaries}

\newcommand{\lmxbs}{low-mass X-ray binaries}

\newcommand{\ztf}{ZTF~J1637$+$49}
\newcommand{\gaia}{Gaia14aae}
\newcommand{\crts}{CRTS~J1122$-$1110}

\newcommand{\ce}{CE}

\newcommand{\popm}{post-orbital-period-minimum}

\newcommand{\bse}{BSE}

\newcommand{\mesa}{MESA}

\defcitealias{Sarkar_2023a}{S23a}

\begin{document}

   \title{Reversing the verdict: Cataclysmic variables could be the dominant progenitors of AM\,CVn binaries after all}


   \titlerunning{CVs could be the dominant progenitors of AM\,CVn binaries}

   \author{Diogo Belloni\inst{1}
          \and
          Matthias R. Schreiber\inst{1,2}
          }

    \authorrunning{D. Belloni \& M. R. Schreiber}

   \institute{Departamento de F\'isica, Universidad T\'ecnica Federico Santa Mar\'ia, Av. España 1680, Valpara\'iso, Chile\\
              \email{diogobellonizorzi@gmail.com}
         \and
             Millenium Nucleus for Planet Formation, Valpara{\'i}so, Chile\\
             \email{matthias.schreiber@usm.cl}
             }

   \date{Received...; accepted ...}

 
  \abstract
   {
   \ams~are potential progenitors of thermonuclear supernovae and strong sources of persistent gravitational wave radiation. For a long time, it has been believed that these systems cannot descend from cataclysmic variables (\cvs), at least not in large numbers, because the initial conditions need to be fine-tuned and, even worse, the resulting surface hydrogen abundance would be high enough to be detected which contradicts a defining feature of \ams.
   }
   {
   Here we show that both claimed weaknesses of the \cv~formation channel for \ams~are model-dependent and rely on poorly constrained assumptions for magnetic braking.
   }
   {
   We performed binary evolution simulations with the \mesa~code for different combinations of post-common-envelope white dwarf and companion masses as well as orbital periods assuming the CARB model for strong magnetic braking.
   }
   {
   We found that \ams~with extremely-low surface hydrogen abundances are one natural outcome of \cv~evolution if the donor star has developed a non-negligible helium core prior to the onset of mass transfer. In this case, after hydrogen envelope exhaustion during \cv~evolution, the donor becomes degenerate and its surface hydrogen abundance substantially drops and becomes undetectable. Our simulations also show that the \cv~formation channel is able to explain the observed \ams~with very low mass and bloated donor stars (Gaia14aae and ZTF~J1637+49).
   }
   {
   \cvs~with evolved donors are likely the progenitors of at least a fraction of \ams.
   }


   \maketitle
%


\section{Introduction}
\label{introduction}


\ams~are ultra-compact (orbital periods in the range of ${\sim5-65}$~minutes) interacting binaries in which a white dwarf accretes helium-dominated material with undetectable amounts of hydrogen from a semi-degenerate or degenerate donor \citep[e.g.][]{Solheim_2010,Ramsay_2018,Green_2018,vanRoestel_2022}.
\ams~~deserve special attention for several reasons.

First, thanks to their short orbital periods \ams~are significant sources of low-frequency gravitational waves to be detected by space-based gravitational wave observatories.
In particular, it is expected that hundreds of such systems will be detectable by the Laser Interferometer Space Antenna (LISA) satellite \citep{AmaroSeoane_2017,AmaroSeoane_2023}, which makes them ideal targets for the performance validation of this satellite \citep{Kupfer_2018}. 
Second, \ams~may produce thermonuclear supernovae \citep{bildstenetal07-1} which are among the most important explosions in the Universe. Third, very recently the first magnetic white dwarfs in \ams~have been identified \citep{Maccarone_2023}, which may provide key constraints on models aiming at explaining the origin and evolution of white dwarf magnetic fields \citep[e.g.][]{Schreiber_2021}.
Unfortunately, despite their relevance for several areas of modern astrophysics, current theories struggle to reliably predict the formation rates and characteristics of \ams.


The formation of an \am~requires at least two episodes of Roche-lobe mass transfer, one to form the accreting white dwarf and another one to form the donor \citep[e.g.][]{BelloniSchreiberREVIEW}.
The resulting formation channels have been summarised by \citet[][]{Solheim_2010}.
In the white dwarf channel \citep[e.g.][]{Deloye_2007,Wong_2021}, a detached double white dwarf is formed first, and due to orbital angular momentum loss through gravitational wave radiation the orbit shrinks until eventually the less massive and bigger helium-core white dwarf fills its Roche lobe which causes the system to become an \am.
In the helium star channel \citep[e.g.][]{Yungelson_2008,Heinke_2013,Wang_2021,BK2021}, a detached binary hosting a white dwarf and a helium star is the direct progenitor of \ams.
If the lifetime of the helium star is sufficiently long, gravitational wave radiation will be strong enough to bring the helium star into contact with its Roche lobe resulting in an \am.
In both these formation channels, the last episode of Roche-lobe mass transfer has to be common-envelope (CE) evolution.

In contrast, in the third formation channel, the so-called cataclysmic variable (\cv) channel \citep[e.g.][]{Tutukov_1985,Podsiadlowski_2003,Liu_2021,Sarkar_2023a}, the last episode of mass transfer leading to the formation of \ams~is dynamically stable.
\ce~evolution leads to the formation of a pre-\cv, that is, a detached binary consisting of a white dwarf with a main sequence (or slightly evolved) companion star, which evolves towards shorter orbital periods mainly due to orbital angular momentum loss through magnetic braking.
When the white dwarf companion fills its Roche lobe, the last episode of mass transfer begins.
If this mass transfer is dynamically stable, the binary becomes a \cv, that is, a white dwarf stably accreting hydrogen-dominated material from a main sequence (or slightly evolved) donor star \citep{Warner_1995_OK,Pala_2020}.
If the donor star is slightly evolved, the donor star slowly loses its hydrogen envelope and may convert into a (semi-)degenerate helium-rich donor during the subsequent evolution.   
It is currently unclear which of the above-mentioned formation channels of \ams~is the dominant one (if any).

Towards a better understanding of the formation of \ams, it is reasonable to take a closer look at recent observational constraints. 
Let's start with the probably closest relatives to \ams, so-called helium \cvs~(\hecvs).
In contrast to \ams, in these systems the accreted helium-rich material still contains enough hydrogen to be detected in their spectra \citep[e.g.][]{Breedt_2012,Green_2020}.
The donor stars in \hecvs~have masses typically below $0.1$\,\Msun. 
The formation of \hecvs~is much less uncertain than that of \ams.
As \hecvs~have donors with a non-negligible surface hydrogen abundance, \ce~evolution as the last episode of mass transfer can be ruled out.
This leaves as the only possible pathway the \cv~channel, that is, \hecvs~can be considered to be direct descendants from \cvs.

This finding is complemented by the recent observational identification of \cvs~currently making the transition from hydrogen-dominated to helium-dominated accretion regimes \citep[e.g.][and references therein]{Green_2020,Lee_2022,Burdge_2022}.
These so-called transitional \cvs~contain a white dwarf accreting helium-rich material from a donor that is more massive (i.e. in the range of ${\sim0.1-0.2}$~\Msun) than those of \hecvs. 
Otherwise these systems are quite similar to \hecvs, that is, they have orbital periods shorter than ${\sim65}$~minutes and show both helium and hydrogen lines. 
Transitional \cvs~and \hecvs~may provide an evolutionary link between \cvs~and \ams. 
Transitional \cvs~are expected to evolve towards shorter orbital periods until the envelopes of their donor stars are hydrogen exhausted and supported by electron degeneracy pressure. From this point on they will evolve towards longer orbital periods, appearing first as \hecvs~and then quickly turn into \ams.

The existence of \hecvs~and transitional \cvs~supports the \cv~channel for the formation of \ams.
Additional evidence for this channel to potentially play a major role has been provided by a recent observational study of \cvs~hosting nuclear evolved donors \citep{ElBadry_2021}. 
This work tripled the number of known \cvs~with evolved donors and estimates a birth rate roughly consistent with that of \ams.
\citet{ElBadry_2021} therefore concluded that the \cv~channel may contribute significantly to the formation of \ams.

As final evidence for the \cv~channel, we mention observational facts that are difficult to explain with the alternative channels for \am~formation. 
The current sample of eclipsing \ams~with reliable mass and radius measurements \cite[][their tab.~8]{vanRoestel_2022} contains two systems, Gaia14aae and ZTF~J1637$+$49, with donors that are significantly bigger than typical \am~donors of the same mass \citep{Green_2020}.
These larger radii resemble those of \hecv~donors and might be naturally explained in the context of the CV channel while it appears rather challenging to understand their formation through the white dwarf and the helium star channels.
In general, the properties of donors belonging to \ams~with orbital periods longer than ${\sim50}$~minutes are difficult to explain with the white dwarf and helium star channels.
This is because irrespective of whether the donor is initially a helium-core white dwarf or a helium star, it will eventually cool and contract during the evolution before reaching such long orbital periods.
To sum up, recently obtained observational constraints support that not only \hecvs~but also transitional \cvs~and \ams~might originate in large numbers from \cvs.

In stark contrast to this recent observational evidence, for a long time the \cv~channel has been considered to be insignificant for the formation of \ams, mainly because of two reasons that are based on the theoretical modelling of the \cv~channel.
First, previous calculations \citep[e.g.][]{Goliasch_2015,Kalomeni_2016} predicted that \ams~are only formed from a very narrow range of the parameter space of initial post-\ce~binaries, which implies that \ams~should be much rarer than they actually are if this channel contributed significantly to their formation. 
It is worth noting that a similar argument has been used in the context of the formation of close detached millisecond pulsars orbiting helium-core white dwarfs. For these systems it also has been found that only a small range of initial parameters can lead to their formation \citep{Istrate_2014}.
Second, previous evolutionary models suggested that the \cv~channel always predicts \ams~to have detectable amounts of hydrogen in their atmospheres 
\citep[e.g.][]{Nelemans_2010}.
Given that \ams~lack hydrogen in their spectra, the classical conclusion was that the \cv~channel should not significantly contribute to the intrinsic population of \ams.

However, the two arguments against the formation of \ams~through the \cv~channel only hold when relatively weak magnetic braking is assumed to drive \cv~evolution. 
Interestingly, in the case of the just mentioned millisecond pulsar binaries, it has recently been shown that orbital angular momentum loss due to sufficiently strong magnetic braking can solve the previously identified fine-tuning problem \citep[e.g.][]{Chen_2021,Deng_2021}.
Given that also for the \cv~channel a dependence of the number of \cvs~with evolved donors turning into \ams~on the assumed magnetic braking prescription has been noted \citep{Podsiadlowski_2003,Liu_2021,Sarkar_2023a}, one might speculate that the fine-tuning problem associated with the \cv~channel for \ams~could also be solved by assuming stronger magnetic braking.
This is simply because the stronger the magnetic braking, the longer the maximum initial post-\ce~binary orbital period that still leads to convergent evolution (i.e. the binary evolves towards shorter orbital periods).
Stronger magnetic braking might also solve the problem of the predicted detectability of hydrogen in \ams~as it may drive convergent \cv~evolution for more nuclear evolved donors, which may lead to more hydrogen-deficient degenerate donors.

We here revisit theoretical predictions of the \cv~channel adopting the CARB model for magnetic braking \citep{CARB}, which leads to sufficiently high orbital angular momentum loss rates.
The main difference between the CARB and previously proposed prescriptions for magnetic braking is that instead of representing a rather empirical prescription, magnetic braking according to the CARB model is obtained in a self-consistent physical way, considering wind mass loss, the dependence of the magnetic field strength on the outer convective zone and the dependence of the Alfv\'en radius on the donor spin.
The main motivation for the development of this magnetic braking model was to explain persistent \lmxbs~and \ucxbs, which is virtually impossible with other prescriptions for magnetic braking \citep{Van_2019,Deng_2021}.
Most importantly for the topic of this paper, the CARB prescription for magnetic braking has also proven to be successful in explaining several types of objects that experience a similar evolution as the one we investigate here \citep[e.g.][]{Soethe_2021}.

We indeed find that sufficiently strong magnetic braking can overcome the frequently mentioned difficulties of the \cv~channel.
We show that if magnetic braking is as strong as that predicted by the CARB model, \cvs~hosting more nuclear evolved donors can still evolve into \ams.
Furthermore, except for AM\,CVn itself, all \ams~with reliable stellar and binary parameters can be explained in the context of the \cv~channel, especially the `problematic' ones hosting oversized donors at long orbital periods, namely \gaia~and \ztf.

%
%
\section{Methodology}
\label{methodology}

We used the \mesa~code \citep[][r15140]{Paxton2011, Paxton2013, Paxton2015, Paxton2018, Paxton2019,Jermyn2023} to compute the evolution of \cvs~and their descendants.
The \mesa~equation of state is a blend of the OPAL \citep{Rogers2002}, SCVH \citep{Saumon1995}, FreeEOS \citep{Irwin2004}, HELM \citep{Timmes2000}, PC \citep{Potekhin2010} and Skye \citep{Jermyn_2021} equations of state.
Nuclear reaction rates are a combination of rates from NACRE \citep{Angulo1999}, JINA REACLIB \citep{Cyburt2010}, plus additional tabulated weak reaction rates \citep{Fuller1985, Oda1994,Langanke2000}.
Screening is included via the prescription of \citet{Chugunov2007} and thermal neutrino loss rates are from \citet{Itoh1996}.
Electron conduction opacities are from \citet{Cassisi2007} and radiative opacities are primarily from OPAL \citep{Iglesias1993,Iglesias1996}, with high-temperature Compton-scattering dominated regime calculated using the equations of \citet{Buchler1976}.
In what follows, we describe in detail the assumptions for stellar and binary evolution as well as the setup of our grid of models and the observational sample we compare the model predictions with.

\subsection{Stellar evolution assumptions}

For low-temperatures the \mesa~code offers two options for the radiative opacities, those from \citet{Ferguson2005} and those from \citet[][]{Freedman2008,Freedman2014}.
As we investigated systems that can reach very low temperatures (${\lesssim1000}$~K), we adopted the latter as they cover significantly lower temperatures than the former.
However, we checked the influence of our selection by also running a set of models with opacities from \citet{Ferguson2005}.
The results of this exercise are described in Sect.~\ref{Problem2}.

We adopted two schemes for the boundary conditions of the atmosphere, both of which are described in \citet[][their sect.~5.3]{Paxton2011}.
Most of the time we assumed the grey Eddington T(tau) relation to calculate the outer boundary conditions.
However, in case the hydrogen envelope was entirely consumed, we switched to non-grey model atmosphere tables, which determine the surface temperature and pressure at the surface optical depth of one.
We adopted a varying opacity consistent with the local temperature and pressure throughout the atmosphere.
We further checked the influence of how opacities are calculated throughout the atmosphere by running a set of models using a uniform opacity that is iterated to be consistent with the final surface temperature and pressure at the base of the atmosphere.
The results of this exercise are also described in Sect.~\ref{Problem2}.

We took into account exponential diffusive overshooting on the main sequence in case the star has a convective core.
More specifically, we incorporated overshooting for stars initially more massive than $1.1$~\Msun~with a smooth transition in the range ${1.1-1.2}$~\Msun.
We assumed that the extent of the overshoot region corresponds to ${0.016~H_{\rm p}}$ \citep[e.g.][]{Schaller_1992,Freytag_1996,Herwig_2000}, with $H_{\rm p}$ being the pressure scale height at the convective boundary.

We further used the nuclear network \texttt{cno$\_$extras.net}, which accounts for carbon-nitrogen-oxygen burning.
Convective regions were treated using the \citet{Henyey_1965} modification of the mixing-length theory assuming that the mixing length is ${2\,H_{\rm p}}$.
For mass loss through winds we adopted the well-known \citet{Reimers_1975} prescription, setting the wind efficiency parameter to $0.5$, which is consistent with metallicity-independent estimates derived from star cluster red giants \citep{McDonald_2015}.
Finally, we assumed solar metallicity (i.e. ${Z=0.02}$) and set all other \mesa~parameters for stellar evolution to their default values.

\subsection{Binary evolution assumptions}

We applied orbital angular momentum loss due to gravitational wave radiation as well as magnetic braking.
For the latter we adopted the CARB prescription as implemented in \mesa~by \citet[][zenodo.3647683\footnote{ \href{https://zenodo.org/record/3647683}{ https://zenodo.org/record/3647683}}]{CARB}, which is given by

\begin{equation}
\begin{multlined}
\dot{J}_{\rm MB} \ = \ 
- \, 2\times10^{-6}
\left(
   \frac{-\dot{M}_{\rm wind}}{\rm g~s^{-1}}
\right)^{-1/3}
\left(
   \frac{R}{\rm cm}
\right)^{14/3}
\left(
   \frac{\Omega}{\Omega_\odot}
\right)^{11/3} \, \times
\\
\left(
    \frac{\tau_{\rm conv} }{\tau_{\odot, \rm conv}}
\right)^{8/3}
\left[
  \left(
      \frac{v_{\rm esc}}{\rm cm~s^{-1}}
  \right)^2\,+\,\frac{2}{K_2^2}\,
  \left(
      \frac{\Omega}{\rm s^{-1}}
  \right)^2
  \left(
      \frac{R}{\rm cm}
  \right)^2
\right]^{-2/3}
\end{multlined}
\label{Eq:MB-CARB}
\end{equation}

\noindent
where $\dot{M}_{\rm wind}$, $R$, $\Omega$, $\tau_{\rm conv}$ are the wind mass-loss rate, radius, spin, and convective turnover timescale of the companion of the white dwarf, respectively.
The convective turnover timescale was calculated by integrating the inverse of the velocity of convective cells, as given by the mixing-length theory, over the radial extent of the convective envelope.
The Sun spin and convective turnover timescale are  ${3\times10^{-6}}$~${\rm s^{-1}}$ and ${2.8\times10^6}$~s, respectively, and ${K_2=0.07}$.
Finally, $v_{\rm esc}$ is the escape velocity.

Magnetic braking was assumed to contribute to the orbital angular momentum loss as long as a non-negligible convective envelope and a non-negligible radiative core are present.
If the mass of the convective envelope decreased to less than two\,percent of the entire donor mass, we reduced the strength of magnetic braking by a factor $e^{1-0.02/q_{\rm conv}}$ \citep{Podsiadlowski_2002}, where $q_{\rm conv}$ is the mass fraction of the convective envelope.
This approach takes into account that donors with a very small convective envelope do not develop strong magnetic fields and will in turn experience little magnetic braking.

In order to calculate evolutionary tracks of \cvs, it is fundamental to consider the different regimes of mass transfer.
Depending on the mass transfer rate and in turn the mass accretion rate onto the white dwarf, hydrogen shell burning on the white dwarf can be stable or unstable \citep[e.g.][]{Shen_2007,Nomoto_2007}.
In case the mass accretion rate is lower than the limit for stable hydrogen burning, nova eruptions occur (unstable hydrogen burning) and mass transfer is highly non-conservative. 
Consistent with simulations of nova cycles \citep[e.g.][]{Yaron_2005}, we assumed that the entire accreted mass is expelled during nova cycles, carrying away the white dwarf specific orbital angular momentum.
If the mass transfer rate exceeds a critical value, we expect stable hydrogen burning in a shell and mass transfer can be highly conservative.
We assumed, in this case, mass transfer to be fully conservative, that is, we assumed that all the accreted mass remains on the white dwarf.
To determine the critical rate separating stable and unstable hydrogen burning, we adopted the criterion from \citet[][their fig.~1]{Wolf_2013}.

If the accretion rate significantly exceeds the value required for stable hydrogen burning, the star evolves into a giant-like structure and the accretion rate is limited by the core mass-luminosity relation, which defines an upper limit for stable hydrogen burning \citep[see][for more details]{Wolf_2013}. 
If the mass transfer rate exceeds this upper limit, we assumed a maximum rate of stable hydrogen burning and that the remaining non-accreted matter will form a red-giant-like envelope that is eventually lost due to strong winds \citep{Hachisu_1996}.
We adopted the criterion from \citet[][their fig.~1]{Wolf_2013} to determine this upper limit.
This procedure is especially important if the mass ratio between the donor and the white dwarf is large enough and/or the donor is already substantially evolved at the onset of mass transfer as in these cases the white dwarf mass can significantly grow during early \cv~evolution.

More accurately, we should adopt different critical values when the transferred material becomes helium-rich.
In this case, the critical values are higher than those adopted in this work \citep[e.g.][]{Wang_2018}, with the limit for stable helium burning being typically higher than ${\sim5\times10^{-6}}$~\Msun~yr$^{-1}$.
Such high rates are never reached in our calculations during accretion of helium-rich material.
While the mass acretion rate can be larger than the smaller critical values that we adopted in this work, this occurs only for a very short amount of time (${\lesssim0.2}$~Myr), which implies that the mass growth we predict during this phase is negligible (${\lesssim3}$\%, corresponding to a few $0.01$~\Msun) and, in turn, that our results are not affected by this.

We set all other \mesa~parameters for binary evolution to their default values.

\subsection{Initial post-CE conditions and post-CE evolution}

We started all our simulations immediately after white dwarf formation, that is, we assume as initial configuration a post-\ce~detached binary consisting of a point-mass white dwarf and a zero-age main-sequence star. 
Throughout this paper we use the term 'initial post-\ce~binary' to refer to these initial conditions of our simulations. 
The initial parameters of the post-\ce~binaries covered white dwarf masses from $0.4$ to $1.0$~\Msun, companion masses from $1.0$ to $1.5$~\Msun, and orbital periods ranging from ${\sim0.25}$ to ${\sim5}$~d.

We assumed that the orbits of post-\ce~binaries circularised during \ce~evolution which is in agreement with observations of large samples of post-\ce~binaries \citep{nebot-gomez-moranetal11-1,Hernandez_2021,Hernandez_2022a,Hernandez_2022b,Zorotovic_2022}.
Finally, the companion was assumed to leave \ce~evolution synchronised with the orbital motion, which means magnetic braking was allowed to extract orbital angular momentum right from the start of the simulations (as long as the other conditions for magnetic braking, such as the presence of a convective envelope and a radiative core, were satisfied).
We then computed the evolution of different \cvs~by varying the initial post-\ce~binary masses and orbital periods.
We are aware of the fact that the evolution prior to \ce~evolution was not taken into account in our simulations.
To include this part of the evolution would require to simulate the formation of the white dwarfs which is beyond the scope of the present investigation but will be addressed in a follow-up publication.

To be considered as a potential progenitor of a \hecv~or an \am~in our simulations, we requested that the donor star has at least a non-negligible helium core (${\geq0.01}$~\Msun) at the onset of mass transfer.
Otherwise, the system is treated as a ``standard'' \cv~and its evolution was no longer taken into account.
Donors with relatively massive helium cores at the onset of mass transfer will likely evolve to \ce~evolution, followed by a merger.
We imposed a critical mass transfer rate of ${10^{-4}}$~\Msun~yr$^{-1}$, above which we assumed that mass transfer is or will become dynamically unstable and consequently discarded systems that reached this value.
Finally, we stopped all simulations when either the donor age exceeded the Hubble time ($\approx13.7$~Gyr), or its mass dropped below $0.01$~\Msun.
The former choice is because this corresponds to the maximum age a system is expected to have, while the latter choice is motivated by the lowest values of the donor mass in \ams~derived from observations.

We considered as a \hecv~or an \am~only those \cvs~that at some point during their evolution reach orbital periods below ${\sim60}$~minutes.
Albeit there is no clear overall distinction between \ams~and \hecvs~that could be applied to all systems \citep[e.g.][]{Nasser_2001,Nagel_2009,Green_2019}, we separated them adopting a critical surface hydrogen mass fraction of $10^{-4}$ based on the upper limits estimated by \citet{Green_2019} for Gaia14aae.
Systems having surface hydrogen abundances lower than that were considered \ams, and systems exceeding this critical value were referred to as \hecvs.

\subsection{Observational samples}
\label{obssample}

To verify whether our evolutionary tracks can reproduce observed samples of \ams~we compared them with observational samples of systems corresponding to different evolutionary stages during post-\ce~binary evolution.

%
For the observational sample of \cvs~we considered systems hosting nuclear evolved donors, which can be easily identified by their much larger donors when compared to an unevolved main-sequence star and from their enhanced \ion{N}{v}/\ion{C}{iv} line flux ratios \citep[but see also the discussion in][]{Sparks_2021}.
We excluded systems for which only estimates or upper limits for the mass ratio are provided in the literature. 
This left us with the systems V1309~Ori \citep{Staude_2001}, EY~Cygni \citep{Echevarria_2007}, AE~Aqr \citep{Echevarria_2008}, HS~0218$+$3229 \citep{RodriguezGil_2009}, KIC~5608384 \citep{Yu_2019}, V1460~Her \citep{Ashley_2020}, CXOGBS~J175553.2--281633 \citep{Gomez_2021}, 1RXH~J082623.6--505741 \citep{Sokolovsky_2022}, and SWIFT~J183221.5--162724	\citep{Beuermann_2022}.
In addition, we included the `terminal' \cvs~discovered by \citep{ElBadry_2021}, which host nuclear evolved but fairly low-mass donors.
In the majority of these systems, mass transfer is still ongoing, as indicated by weak emission lines, eclipses of the donor by an accretion disk, and irregular variations in brightness.
The remaining systems have recently detached, with donors being proto-white dwarfs that nearly fill their Roche lobes.

%
Among all observed \ams, we consider only those that are either eclipsing or have mass ratios derived from spectroscopy, as we consider the derived parameters in these cases to be more reliable.
The well-characterized eclipsing systems that we include are YZ~LMi \citep{Copperwheat_2011}, Gaia14aae \citep{Green_2018}, and the five Zwicky Transient Facility (ZTF) systems discovered by \citet{vanRoestel_2022}, namely ZTF~J1637$+$49, ZTF~J0003$+$14, ZTF~J0220$+$21, ZTF~J2252--05, and ZTF~J0407--00.
Systems with mass ratios derived from spectroscopy are GP~Com \citep{Marsh_1999}, SDSS~J1240--0159 \citep{Roelofs_2005}, AM~CVn itself \citep{Roelofs_2006}, and V396~Hya \citep{Kupfer_2016}.
For the last four systems, only the orbital period and mass ratio has been determined from observations. We therefore used the estimates by \citet{Green_2018b} for the radii and masses of the donors.
There are several other observed \ams~with mass ratio estimated from the superhump excess–-mass ratio relation.
Even though this procedure seems to work reasonably well for \cvs, it is not yet clear whether this approach leads to accurate estimates or not when applied to \ams~\citep[e.g.][]{Roelofs_2006,Green_2018b}.
For this reason, we did not consider these systems.

%
For none of the currently known \hecvs~parameters have been derived from eclipses or spectroscopy, as highlighted by \citet{Green_2020}.
For these systems, the mass ratio was derived from the superhump excess--mass ratio relation.
To avoid excluding these important systems from our study, we consider them here but draw the readers' attention to the fact that the estimated properties are not necessarily reliable.
The systems we consider are CRTS~J1740$+$4147 \citep{Imada_2018}, CRTS~J1122--1110 \citep{Breedt_2012} and CRTS~J1111$+$5712 \citep{Littlefield_2013}.
For these three systems, we took the values of the donor mass and the donor radius estimated by \citet{Green_2020}.

%
Finally, regarding transitional \cvs, there is only one that is eclipsing with measured parameters, that is, ZTF~J1813+4251 discovered by \citet{Burdge_2022}.
The remaining systems have properties estimated from superhump excess, except for V485~Cen, whose mass ratio was derived from spectroscopy \citep{Augusteijn_1996}.
To include more systems of this class, we considered not only ZTF~J1813+4251 and V485~Cen but also the systems CRTS~J1028--0819 \citep{Woudt_2012} and EI~Psc \citep{Thorstensen_2002}.
We used the parameters derived by \citet{Burdge_2022} for ZTF~J1813+4251, while we took the masses and radii of the donors estimated by \citet{Green_2020} for the other systems.

\section{Cataclysmic variable evolution revisited}
\label{resultsCVEVOLUTION}

We start the presentation of our results by revisiting \cv~evolution.
In case the donor star is not evolved and assuming the often used prescription of disrupted magnetic braking based on \citet{RVJ} one obtains standard evolutionary tracks for \cv~evolution \citep[e.g.][]{Knigge_2011_OK,Schreiber_2016,Belloni_2018b,Belloni_2020a,Pala_2020,Pala_2022}, an example of which is given by the long-dashed red lines in Fig.~\ref{FigE} \citep[extracted from][]{BelloniSchreiberREVIEW}.
For this evolutionary track we assumed a post-\ce~orbital period, white dwarf mass and companion mass of $0.5$~d, $0.8$ and $0.8$~\Msun, respectively.

In brief, following the onset of mass transfer, the partially convective donor is driven out of thermal equilibrium and expands in response to mass loss due to sufficiently high orbital angular momentum loss through magnetic braking.
The donor is nevertheless able to reach thermal equilibrium, but with a radius that is larger than expected for its mass.
The \cv~remains in a semi-detached configuration until the donor becomes fully convective, at an orbital period of ${\approx3}$~h.
At this moment, magnetic braking becomes significantly weaker \citep{schreiberetal10-1,Zorotovic_2016}, probably because of a rise in the donor magnetic field complexity \citep[e.g.][]{TaamSpruit_1989,Garraffo_2018}.
As a consequence of the strongly reduced orbital angular momentum loss rate, the donor can relax towards its thermal equilibrium, shrinks to its normal size and loses contact with its Roche lobe, and the system becomes a detached binary.
After the remaining orbital angular momentum loss mechanisms have sufficiently shortened the orbital period, mass transfer starts again at a lower rate at an orbital period of ${\approx2}$~h.
The detached phase in the period range of ${\approx2-3}$~h is the standard explanation for the observed orbital period gap.

The situation changes for evolved donor stars.
In this case, the detached phase producing the orbital period gap is not expected as the donor does not become fully convective.
However, \cv~evolution with evolved donors may include a detached phase caused by a different effect. 
If the donor is sufficiently evolved, its envelope might become radiative which should suppress magnetic braking. As the evolutionary pathways of \cvs~with evolved donors have implications for the emerging \ams, we now describe in detail both possibilities,
that is, \cv~evolution and \am~formation with and without a detached phase.

\subsection{Evolution with a detached phase}
\label{resultsCVEVOLUTIONwithGAP}

A standard condition required for efficient magnetic braking is the existence of a non-negligible convective envelope, a fact that is taken into account in the CARB model.
Therefore, magnetic braking can be suppressed during the evolution of \cvs~with evolved donors in case the convective envelope becomes radiative.
This occurs if the effective temperature of the donor exceeds a critical value \citep[e.g.][]{Belczynski_2008}, which corresponds to the convective/radiative envelope boundary for subgiants. 
This boundary is in general very broad (${\sim5000-8000}$~K) and the exact value varies from system to system, depending on stellar properties such as metallicity, mass, structure, outer boundary conditions and chemical profile.
We provide more details about the donor effective temperature evolution in Sect.~\ref{resultsCVEVOLUTIONcontraction}.

\begin{figure}
\begin{center}
\includegraphics[width=0.99\linewidth]{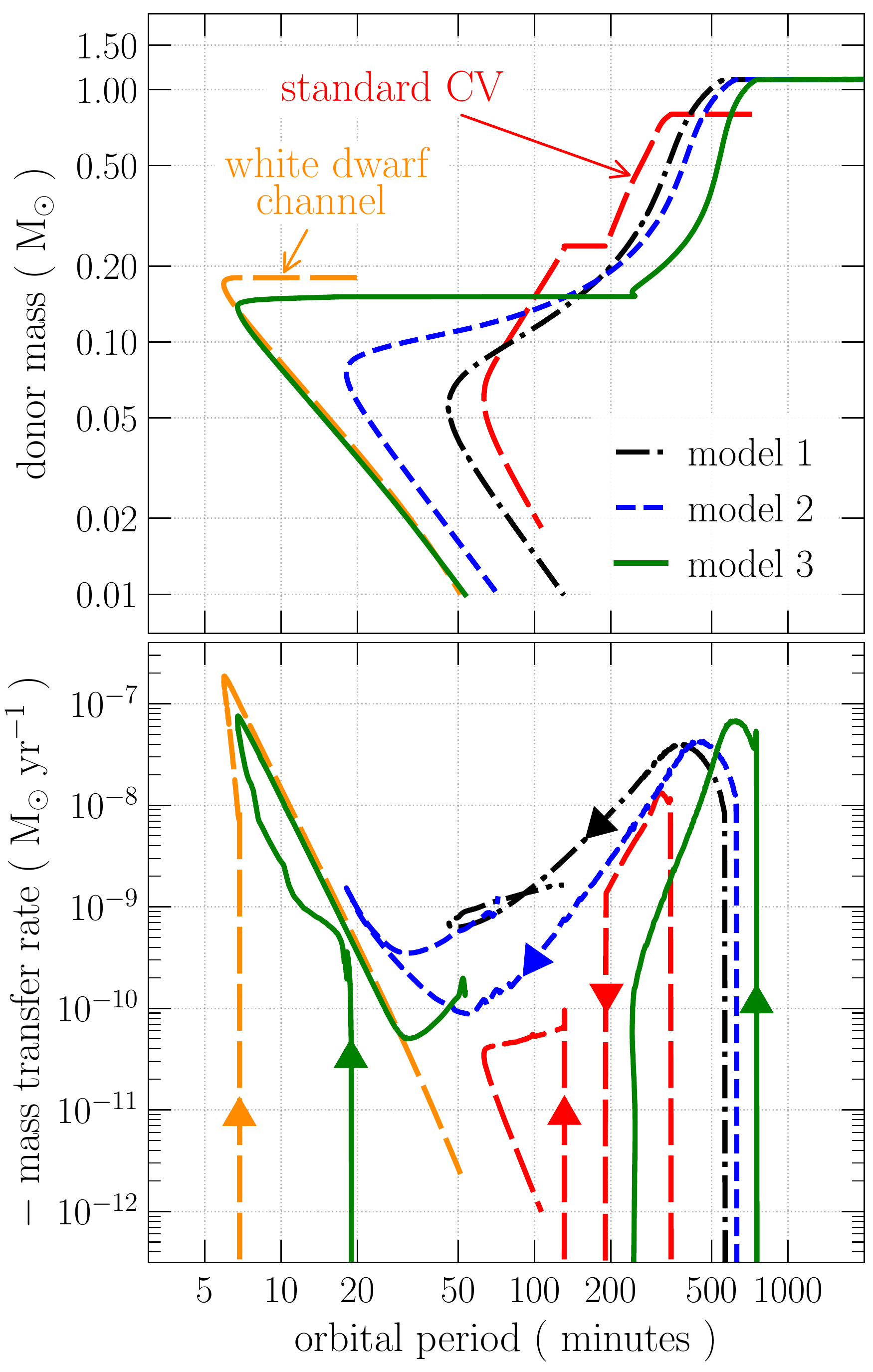}
\end{center}
\caption{
Evolution of donor mass (top panel) and mass transfer rate (bottom panel) as a function of orbital period obtained adopting the CARB model for magnetic braking.
We show the evolution of three illustrative \cvs~with initial post-\ce~white dwarf and donor star masses of $0.8$ and $1.1$~\Msun, respectively. The initial post-\ce~orbital periods and donor helium core mass at the onset of mass transfer are different in each evolutionary track and are given by the following values:
model~1: $2.20$~d and ${\approx0.043}$~\Msun~(dot-dashed black lines),
model~2: $2.34$~d and ${\approx0.068}$~\Msun~ (short-dashed blue lines), and
model~3: $2.60$~d and ${\approx0.105}$~\Msun~ (solid green lines).
%
%
On top of that, we also show an example of standard \cv~evolutionary sequence \citep[long-dashed red lines, from][]{BelloniSchreiberREVIEW} and an example of evolutionary sequence for the white dwarf channel \citep[long-dashed orange lines, from][]{Wong_2021}.
The arrows in the bottom panel were added for clarity and indicate the direction of the evolution.
The tracks for which the orbital period minimum is below $20$ minutes correspond to \cv~evolution leading to the formation of \ams.
Model~3 represents a mixed channel in which a \cv~leads to a close detached double white dwarf binary that latter becomes an \am. The final evolution of this system is similar to the white dwarf channel for \am~formation.
See text for more details.
}
\label{FigE}
\end{figure}

The transition from convective to radiative envelope occurs for sufficiently hot and nuclear evolved donors. 
An example for this evolution is given by model~3 in Fig.~\ref{FigE}, which has been calculated assuming values $2.60$~d, $0.8$ and $1.1$~\Msun~for the initial post-\ce~orbital period, white dwarf mass and donor mass,  respectively.
At the onset of mass transfer, which takes place $6.72$~Gyr after \ce~evolution, the mass of the helium core of the donor is ${\approx0.105}$~\Msun~and the convective envelope of the donor corresponds to ${\approx13}$\,\% of its mass.
During early \cv~evolution, the fractional convective envelope mass of the donor increases, reaching a maximum of ${\approx24}$\,\% when the donor mass is ${\approx0.49}$~\Msun, which happens ${\approx11}$~Myr after the onset of mass transfer.
From this point on, the size of the convective envelope decreases as the donor mass decreases and consequently magnetic braking also drops which leads to a fast decrease in the mass transfer rate.

The convective envelope vanishes when the donor effective temperature is ${\approx8029}$~K at a donor mass of ${\approx0.151}$~\Msun~and an orbital period of ${\approx3.95}$~h, which happens ${\approx987}$~Myr after the onset of mass transfer.
At this point, the mass of the helium core of the donor is ${\approx0.128}$~\Msun, which means it significantly increased during \cv~evolution (by ${\approx20}$\,\%).
Magnetic braking is then completely suppressed and further evolution is driven mainly by orbital angular momentum loss due to the emission of gravitational waves.
As this orbital angular momentum loss mechanism is much weaker than magnetic braking, the donor is able to relax to its normal size causing it to detach from its Roche lobe and the system becomes a detached binary through very much the same mechanism that is supposed to cause the orbital period gap of \cvs.
Mass transfer resumes ${\approx3.26}$~Gyr after the donor star detached from its Roche-lobe at a very short orbital period of ${\approx19}$~minutes. 
At this point, the donor mass is ${\approx0.149}$~\Msun~and its helium core mass is ${\approx0.139}$~\Msun.

This detached phase is observationally supported by the results achieved by \citet{ElBadry_2021} who found a group of \cvs~hosting highly consumed subgiant/red giant donors that either just have entered the detached phase or might be very close to it.
Unlike \cvs~harbouring unevolved donors on the main sequence for which the detached phase starts and ends for all systems more or less at the same orbital period, for \cvs~with nuclear evolved donors the width of the detached phase is strongly dependent on the donor properties at the onset of mass transfer, mainly the helium core mass (compare the long-dashed red and solid green tracks in Fig.~\ref{FigE}).
We note that this detached phase has been used not only to explain observations of white dwarf binaries but also close detached millisecond pulsar binaries, which originate from low-mass X-ray binary evolution \citep[e.g.][]{Istrate_2014,Chen_2021,Soethe_2021}.

The longest phases of detached evolution occur when the mass of the helium core of the donor is relatively high (i.e. ${\gtrsim0.14}$~\Msun) at the moment magnetic braking becomes inefficient.
For these systems, mass transfer only resumes at around the orbital period minimum.
However, although the mass of the hydrogen envelope is extremely low (i.e. ${\lesssim0.01}$~\Msun), it always has to be consumed before the system finally reaches the orbital period minimum.
In these cases, the \am~formation pathway can be considered a mixture of the \cv~channel and the white dwarf channel. 
This is because after the mass transfer ceases, the binary simply consists of a white dwarf paired with a newly born extremely-low-mass white dwarf.
Then, as in the white dwarf channel the orbit shrinks until the moment the extremely-low-mass white dwarf fills its Roche lobe at the onset of \am~evolution.
In other words, an \am~can form through the white dwarf channel even when the donor was formed via an episode of dynamically stable mass transfer in the \cv~channel. 
Figure~\ref{FigE} shows that the final evolution of detached \cvs~with evolved donors can be very similar to the evolutionary sequence for the white dwarf channel computed by \citet[][zenodo.5532940\footnote{ \href{https://zenodo.org/record/5532940}{ https://zenodo.org/record/5532940}}]{Wong_2021}, corresponding to their model with initial accretor mass, donor mass and donor specific central entropy of, respectively, $0.75$~\Msun, $0.18$~\Msun~and $3.07$~$N_{\rm A}k_{\rm B}$, where $N_{\rm A}$ is Avogadro's number and $k_{\rm B}$ is the Boltzmann's constant.

However, there is an important difference between the donors resulting from \cv~evolution and those having formed through \ce~evolution. 
In the white dwarf channel, the donor entropy is regulated by the double white dwarf formation history and initial post-\ce~properties \citep[e.g.][]{Deloye_2007}.
In particular, for a given combination of initial post-\ce~accretor and donor masses the shorter the initial post-\ce~orbital period the higher the donor entropy at the onset of mass transfer.
This holds irrespective of whether the donor was formed first or second during the pathway from a zero-age main-sequence binary to a double white dwarf binary.
On the other hand, in the \cv~channel, when mass transfer resumes following the detached phase, the donor entropy is entirely determined by the helium core mass at the onset of \cv~evolution.
This is because the upper and lower edges of the detached phase depend on the helium core mass at the onset of \cv~evolution.
More specifically, the higher the helium core mass at the onset of \cv~evolution, the longer the orbital period corresponding to the upper edge of the detached phase, and in turn, the longer the time it takes gravitational wave radiation to bring the donor into contact with its Roche lobe again.
This then implies that the higher the helium core mass at the onset of \cv~evolution, the lower the donor entropy and the more degenerate the donor when mass transfer resumes after the detached phase.

Another relevant difference is connected with the importance of magnetic braking when the donor star develops a non-negligible convective envelope during \popm~evolution.
In the classical white dwarf channel, the main driver of \am~evolution is always gravitational wave radiation.
However, our model also allows a degenerate donor to lose orbital angular momentum owing to magnetic braking as long as it develops a non-negligible convective envelope.
In our scenario, magnetic braking typically becomes relevant again,
comparable to or stronger than gravitational wave radiation, as soon as the convective turnover timescale is sufficiently long.
The transition from gravitational-wave-radiation-driven evolution to magnetic-braking-driven evolution takes place at different masses and different orbital periods, depending on the detailed structure of the donor after having passed the orbital period minimum.
In general, the lower the helium core mass at the onset of mass transfer, the higher the donor mass and the shorter the orbital period at which magnetic braking starts to dominate again, which can be seen by comparing models~2 and 3 in Fig.~\ref{FigE}.
Our model therefore predicts different degrees of bloating for donor stars of \ams, depending on their detailed structure at the onset of mass transfer.

Under certain conditions, magnetic braking may also turn on before the orbital period minimum, which happens for model~3 in Fig.~\ref{FigE}.
In general, the higher the helium core mass at the onset of the detached phase, the shorter the orbital period at which mass transfer resumes following the detached phase.
This is a consequence of the relation between the helium core mass and structure of the donor, especially the radius.
For \cvs~with donors having at the onset of the detached phase a moderate helium core mass (${\sim0.12-0.14}$~\Msun), mass transfer resumes at an orbital period significantly longer than the orbital period minimum.
In this case the donor may cross the convective/radiative envelope boundary well before the orbital period minimum but magnetic braking is much weaker than gravitational wave radiation at this point of the evolution.
Irrespective of whether a negligible fraction of angular momentum loss comes form magnetic braking the system evolves towards shorter orbital periods until the hydrogen envelope is entirely consumed.
At this point, the system bounces and subsequently evolves towards longer orbital periods hosting a degenerate donor.

\subsection{Evolution without a detached phase}
\label{resultsCVEVOLUTIONwithoutGAP}

A detached phase can be entirely suppressed when the donor is not sufficiently evolved, that is, when the helium core mass at the onset of mass transfer is ${\lesssim0.1}$~\Msun.
This is illustrated in Fig.~\ref{FigE} by model~2 which was calculated assuming an initial post-\ce~orbital period of $2.34$\,d and a white dwarf mass and donor mass of $0.8$ and $1.1$~\Msun, respectively.
In these cases, the donor never becomes hot enough to cross the convective/radiative envelope boundary, never detaches from its Roche lobe, and magnetic braking only becomes weaker as the donor mass decreases but never ceases. 
At some point, however, when the orbital period is sufficiently short, orbital angular momentum loss caused by the emission of gravitational waves becomes stronger than magnetic braking and drives the evolution towards the orbital period minimum.

For model~2 in Fig.~\ref{FigE}, this transition occurs when the donor mass is ${\approx0.11}$~\Msun~at an orbital period of ${\approx55}$~minutes . As orbital angular momentum loss through magnetic braking further decreases but that through gravitational wave radiation increases towards shorter orbital periods, this transition corresponds to a minimum of the mass transfer rate.
During the subsequent evolution, the mass transfer rate reaches a local maximum at the orbital period minimum (${\approx18}$~minutes for model~2) and afterwards decreases as the orbit expands due to the donors response to mass loss.
As the donor further loses mass and changes its structure (the fractional convective envelope increases), at some point magnetic braking becomes stronger than gravitational wave radiation again, just as discussed in the previous section.
The moment at which magnetic braking and gravitational wave radiation become comparable corresponds to a local minimum of the mass transfer rate because magnetic braking becomes stronger as the donor mass further decreases (and the orbital period increases).
For this reason, \ams~predicted by the \cv~channel that follow this pathway (i.e. without a detached phase) have donors with a relatively high central entropy that are larger than expected for their mass.
We will discuss in more detail the entropy and bloating of donors for cases lacking a detached phase in Sect.~\ref{Problem1MB}.

The last evolutionary pathway we discuss is related to \cvs~giving birth to \hecvs~without transitioning into \ams~at donor masses ${\gtrsim0.01}$~\Msun, which is illustrated by model~1 in Fig.~\ref{FigE}.
This evolution corresponds to the case of the least nuclear evolved donors, that is, donors with low helium core masses (${\lesssim0.04}$~\Msun) at the onset of mass transfer.
The evolution in these cases is similar to that of model~2, except that magnetic braking never becomes weaker than gravitational wave radiation.
Throughout the evolution, the fractional mass of the convective envelope of the donor stays at a level of at least  ${\sim20}$\,\%, and for this reason magnetic braking remains efficient.
This evolution also corresponds to the case of the highest donor entropies when the hydrogen envelope vanishes, leading to the largest donors with the lowest degree of degeneracy and longest orbital period minimum (${\approx45}$~minutes for model~1).

\begin{figure*}
\begin{center}
\includegraphics[width=0.99\linewidth]{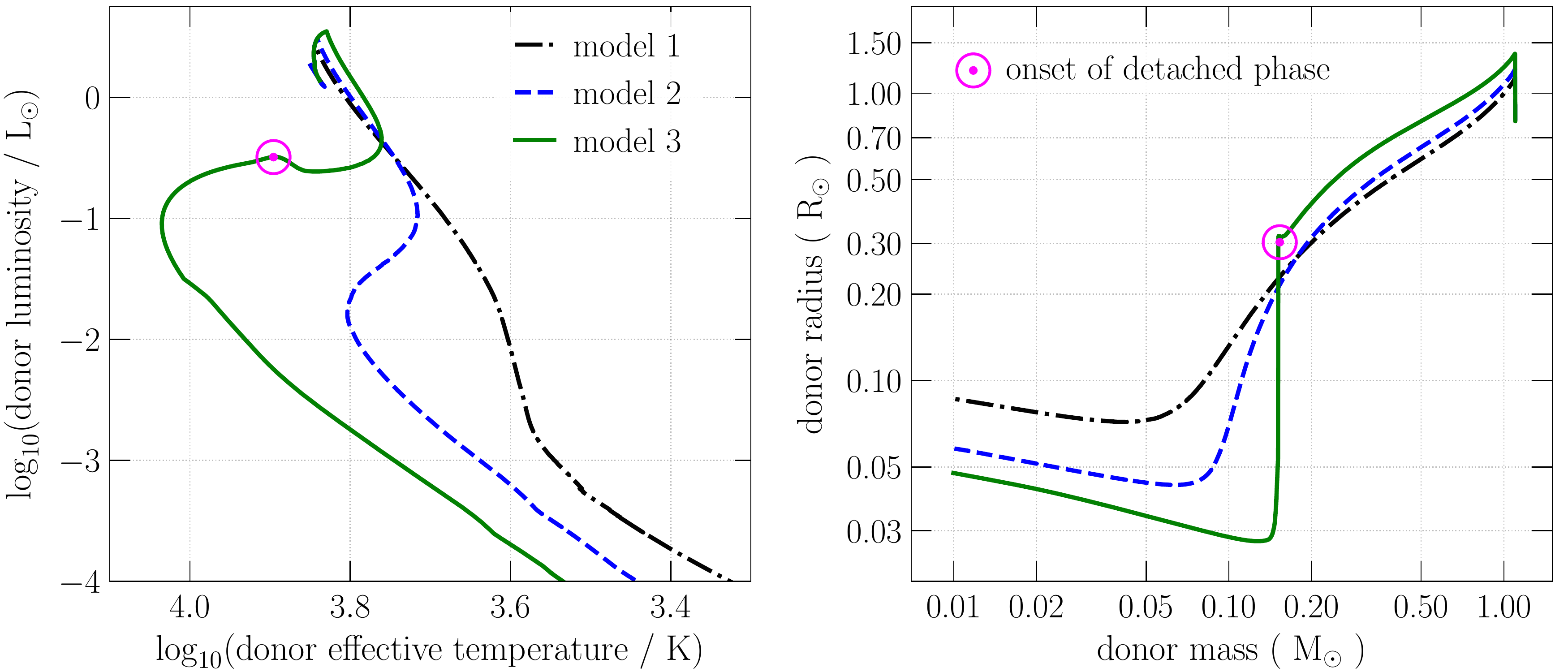}
\end{center}
\caption{
Hertzsprung–Russell diagram (left panel) and mass--radius evolution (right panel) of the donors of the three models shown in Fig.~\ref{FigE}, namely model~1 (dot-dashed black lines), model~2 (short-dashed blue lines) and model~3 (solid green lines).
Following the onset of mass transfer, the donor luminosity, effective temperature and radius decrease.
The radius of donors with relatively massive helium cores (${\gtrsim0.05}$~\Msun) at the onset of mass transfer eventually starts to decrease faster during the evolution.
In case the effective temperature reaches sufficiently high values, the convective envelope vanishes which causes magnetic braking to become inefficient and the donor to detach from its Roche-lobe (illustrated by the magenta circles for model~3).
The more evolved the donor at the onset of mass transfer, the higher the donor mass at the moment the radius starts to decrease faster and the higher the maximum effective temperature reached during the evolution.
}
\label{FigHR}
\end{figure*}

\subsection{Rapid decrease in the donor radius before the orbital period minimum}
\label{resultsCVEVOLUTIONcontraction}

The three models shown in Fig.~\ref{FigE} illustrate the impact of having a nuclear evolved donor at the onset of mass transfer on \cv~evolution.
When the donor mass is ${\gtrsim0.15-0.20}$~\Msun, the more evolved the donor at the onset of mass transfer, the larger its radius during the evolution, which implies that more evolved donors fill their Roche lobes at longer orbital periods.
This is different to what is predicted for \cvs~hosting unevolved main-sequence donor stars at the onset of mass transfer, which follow a very tight evolutionary path as evidenced by the small scatter in the radius--mass relation of these donors \citep[e.g.][]{McAllister_2019}.

At donor masses ${\lesssim0.15-0.20}$~\Msun, the previously mentioned trend is reversed, that is, the more evolved the donor at the onset of mass transfer, the smaller the radius during the late evolution.
This is a well-known feature during the evolution of interacting binaries hosting nuclear evolved donors \citep[e.g.][]{Tutukov_1985,Fedorova_1989,Ergma_1996,Podsiadlowski_2002,Nelson_2004,Lin_2011,Kalomeni_2016,Van_2019,CARB,Deng_2021,Chen_2021,ElBadry_2021,DAntona_2022,Gossage_2023,Yamaguchi_2023}.

How fast the radius decrease and the mass at which this starts to occur strongly depends on how nuclear evolved the donor was at the onset of mass transfer.
The higher the mass of the helium core of the donor at the onset of mass transfer, the faster the radius decreases and the higher the mass at which this starts to happen.
The donor effective temperature at some point stops to increase and starts again to decrease.
The maximum effective temperature reached during the evolution also strongly depends on how nuclear evolved the donor was at the onset of mass transfer.
The higher the mass of the helium core of the donor at the onset of mass transfer, the higher the maximum effective temperature that is reached during the rapid decrease in radius.

Figure~\ref{FigHR} illustrates the above described processes in a Hertzsprung–Russell and mass--radius diagram.
As the donor in model~1 is not sufficiently evolved at the onset of mass transfer, its luminosity and effective temperature monotonically decreases throughout the evolution.
On the other hand, models~2 and 3 have sufficiently massive helium cores at the onset of mass transfer and for this reason at some point their radii start to decrease faster and the thin hydrogen envelope heats up.
The effective temperatures of their donors then switch from slowly decreasing to significantly increasing.
As mentioned in Sect.~\ref{resultsCVEVOLUTIONwithGAP}, the donor in model~3 is sufficiently evolved for its effective temperature to cross the convective/radiative envelope boundary, which causes magnetic braking to become inefficient and the system to enter a detached phase.

The above-discussed processes have an impact on the evolution of the donor mass with the orbital period.
As mentioned earlier, more nuclear evolved donors fill their Roche lobes at longer orbital periods, for masses ${\gtrsim0.15-0.20}$~\Msun.
However, due to the rapid decrease of the radius of those donors that are sufficiently evolved, at masses ${\lesssim0.15-0.20}$~\Msun, donors of the same mass are smaller when they originate from more nuclear evolved donors.
Thus, the more nuclear evolved the donor at the onset of mass transfer, the smaller its radius at these masses.
This implies that donors originating from more evolved donors fill their Roche lobes at shorter orbital periods at masses ${\lesssim0.15-0.2}$~\Msun~as shown in Fig.~\ref{FigE}.

\begin{figure}
\begin{center}
\includegraphics[width=0.99\linewidth]{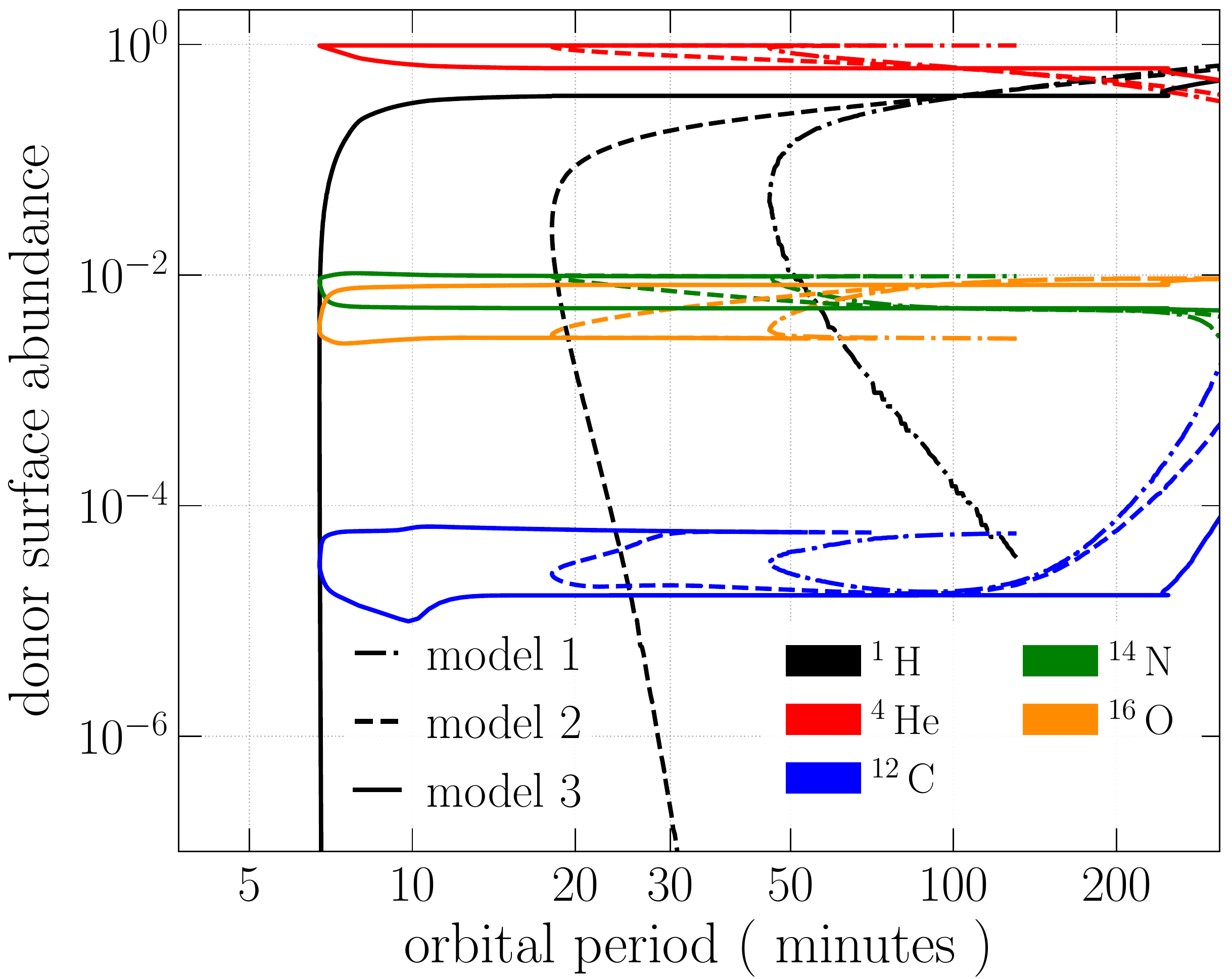}
\end{center}
\caption{
Evolution around the orbital period minimum of the donor surface abundances (colour coded by elements) for the three models shown in Fig.~\ref{FigE}, namely model~1 (dot-dashed lines), model~2 (short-dashed lines) and model~3 (solid lines).
The hydrogen abundance quickly vanishes for model~3, which is similar to the white dwarf channel, when the system reaches the orbital period minimum.
On the other hand, for model~2, the hydrogen drops to an undetectable level during early \popm~evolution (between ${\approx20}$ and ${\approx30}$~minutes).
Finally, the surface hydrogen abundance in model~1 remains high enough to be detected throughout its evolution.
Meanwhile helium abundance is enhanced as the three models approach the orbital period minimum, and together with the other elements, quickly evolves towards equilibrium afterwards.
That said, model~1 is a typical example of how \cvs~evolve to \hecvs.
More importantly, models~2 and 3 clearly illustrate that hydrogen is not necessarily visible in \cv~descendants, irrespective of the present-day orbital period.
In addition, the surface abundances predicted for the two pathways leading to \ams, that is, with (model~3) and without (model~2) a detached phase, are rather similar after hydrogen vanishes suggesting that it is very hard to distinguish between the white dwarf channel and the \cv~channel with respect to abundance ratios.
}
\label{FigX}
\end{figure}

\begin{figure*}
\begin{center}
\includegraphics[width=0.99\linewidth]{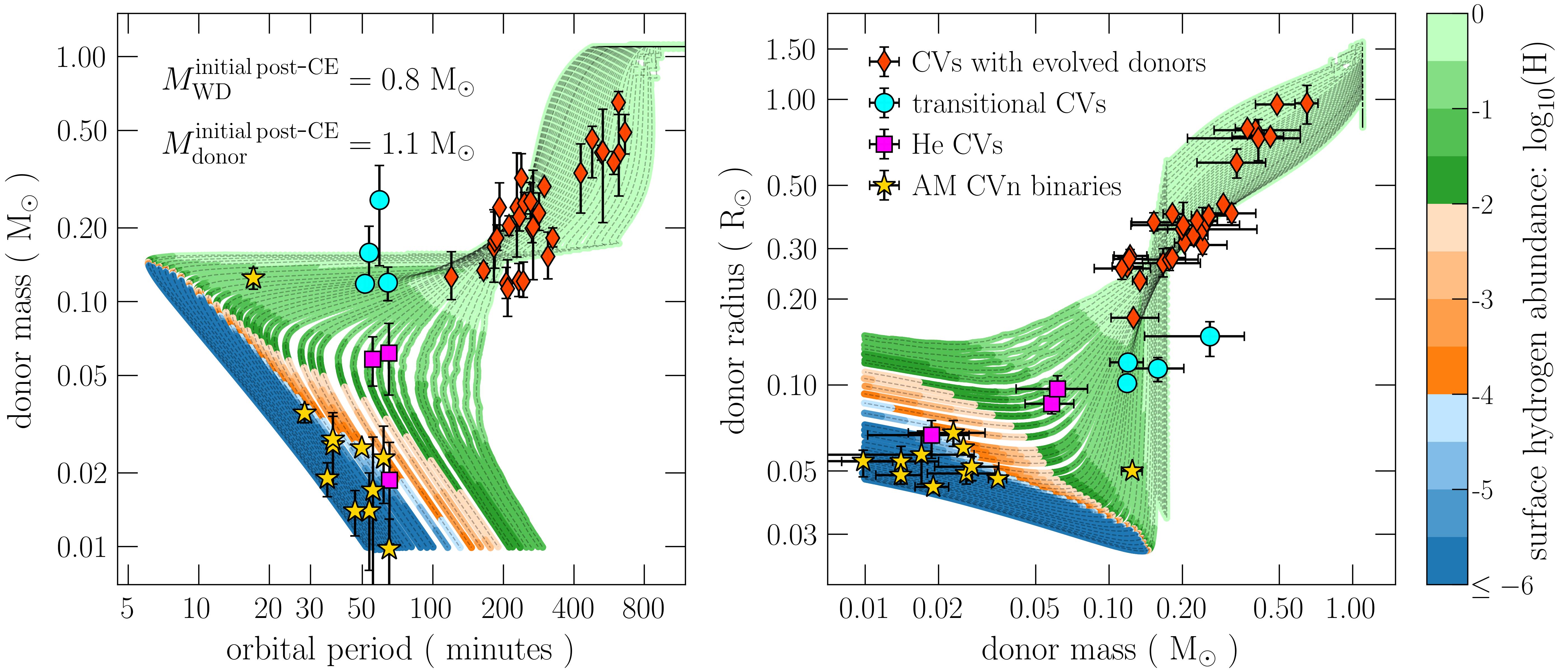}
\end{center}
\caption{
Convergent \cv~evolution when the CARB model for magnetic braking is adopted for different initial post-\ce~binaries.
The colour represents the log-scaled surface hydrogen abundance of the donor.
Systems can be considered as \ams~if they reach shades of blue.
Related objects, that is, \cvs, transitional \cvs, and \hecvs~are indicated by shades of other colours.
While the assumed initial post-\ce~orbital periods cover values from $1.98$ to $3.00$~d, we fixed the initial post-\ce~white dwarf mass to $0.8$~\Msun~and the initial post-\ce~companion mass to $1.1$~\Msun.
The evolution of the surface hydrogen abundance is strongly dependent on the mass of the helium core at the onset of mass transfer: the more massive the helium core, the lower the surface hydrogen abundance at a given orbital period during \popm~evolution.
\cvs~with donors having initially less massive helium cores bounce at longer orbital periods, leading to more massive and larger donors during \am~evolution.
Along with the evolutionary tracks we included the following observed systems: \cvs~harbouring nuclear evolved donors (red diamonds), transitional \cvs~(cyan circles), \hecvs~(magenta squares), and \ams~(yellow stars), as described in Sect.~\ref{obssample}.
Except for AM\,CVn itself, the one with largest donor mass among \ams, the characteristics of all \ams~can be explained by the \cv~channel, which means that the lack of hydrogen lines in the optical spectra of \ams~cannot be easily used as an argument against the \cv~formation channel.
See Sect.~\ref{Problem1} for more details.
}
\label{FigT}
\end{figure*}

\subsection{Surface chemical abundances of the donor}
\label{resultsCVEVOLUTIONsurfaceabundances}

As outlined in the introduction, the predicted detectability of hydrogen is usually considered a major problem of the \cv~formation channel. 
We therefore show in Fig.~\ref{FigX} the evolution of the surface abundances of the donor stars around the orbital period minimum produced by the three models shown in Fig.~\ref{FigE}.
The donors in \ams~that evolved through a detached phase (e.g. model~3) very quickly (${\lesssim0.1}$~Myr) lose their hydrogen while bouncing at the orbital period minimum.
Other elements also reach equilibrium at the orbital period minimum.
There is thus no doubt that \cvs~can give birth to \ams~if their evolutionary pathway includes a detached phase, as no hydrogen would be detected in such systems during \popm~evolution.
A much more detailed discussion of the detectability of hydrogen in the systems predicted by our model is provided in Sect.~\ref{Problem1}.

One may wonder whether the evolutionary pathway without a detached phase can lead to the formation of \ams, that is, whether hydrogen can eventually drop below detectable levels during \popm~evolution.
This can indeed happen as illustrated in Fig.~\ref{FigX}.
As these systems approach the orbital period minimum, the surface hydrogen abundance slightly decreases, while the surface helium abundance slightly increases and the surface abundances of other elements remain roughly constant.
After the system bounces, the surface hydrogen abundance quickly (${\lesssim50}$~Myr) drops during early \popm~evolution and can become undetectable.
For model~2, this happens for orbital periods between ${\approx18}$ and ${\approx25}$~minutes and donor masses between ${\approx0.075}$ and ${\approx0.04}$~\Msun.
For orbital periods ${\gtrsim18}$~minutes and donor masses ${\gtrsim0.04}$~\Msun, this system is a \hecv, while for orbital periods ${\gtrsim25}$~minutes and donor masses ${\lesssim0.04}$~\Msun, it is an \am.
Therefore, for model~2, at an orbital period of ${\approx25}$~minutes and a donor mass of ${\approx0.04}$~\Msun~the system converted from a \hecv~to an \am.
We will discuss this transition in more detail in Sect.~\ref{Problem1TR}.

The surface abundances of elements other than hydrogen remain in equilibrium, except for carbon which is initially enhanced and evolves towards equilibrium as hydrogen diminishes.
Interestingly, the predicted surface abundances are very similar to those produced by evolutionary pathways that include a detached phase.
In particular, we predict ratios between nitrogen and oxygen in the range ${\sim1-7}$, between nitrogen and carbon in the range ${\sim113-220}$, and between oxygen and carbon in the range ${\sim16-130}$.
This indicates that it will be hard to distinguish between the detached \cv~channel and the typical \cv~channel based solely on the chemical composition.

Finally, when the donor is only slightly nuclear evolved at the onset of mass transfer, the overall surface hydrogen abundance remains high throughout the evolution as shown in Fig.~\ref{FigX}.
For model~1, the surface abundances evolve similar to those of model~2, except for hydrogen.
The timescale for hydrogen disappearance is much longer than for the other two models.
In other words, albeit the surface hydrogen abundance drops during \popm~evolution, this effect is just not strong enough for hydrogen to become undetectable.
Such systems therefore appear as \hecvs~for most of their \popm~evolution, likely becoming \ams~only when their donors reach very low masses (comparable to that of planets).

\section{Towards solving the problems of the cataclysmic variable channel}
\label{resultsSOLUTION}

After this brief overview on how \cvs~can evolve into \ams~if the CARB model for magnetic braking is assumed, we turn now to discussing the frequently mentioned problems of the \cv~formation channel that were highlighted in the introduction. We start with the absence of hydrogen lines in the observed spectra of \ams, and subsequently address the fine-tuning problem.

\subsection{Solving the first problem: lack of visible hydrogen}
\label{Problem1}

To evaluate the conditions under which the \cv~channel leads to the formation of \ams~with undetectable amounts of hydrogen, we calculated evolutionary tracks for fixed masses of the initial post-\ce~binaries, that is, a white dwarf mass of $0.8$~\Msun~and a companion of mass of $1.1$~\Msun, and varied the initial post-\ce~orbital period from $1.98$ to $3.00$~d.
The resulting evolutionary tracks are shown in Fig.~\ref{FigT}. 
In both panels, the colours indicate the log-scaled surface hydrogen abundance during the evolution. 
On top of the evolutionary tracks we included the observational samples, as described in Sect.~\ref{obssample}, corresponding to \cvs~hosting nuclear evolved donors, transitional \cvs, \hecvs, and \ams.

The evolution predicted for \cvs~with evolved donors at orbital periods longer than $\sim80$ minutes 
agrees well with the location of the observed \cvs~harbouring nuclear evolved donors. 
The predicted orbital period during this early evolution as well as the donor radii depend strongly on the donor structure at the onset of mass transfer, in particular on its radius and helium core mass.
The shorter the initial post-\ce~orbital period, the sooner the donor fills its Roche lobe, and consequently the smaller its radius and the lower its helium core mass.
For this reason the orbital periods at the onset of mass transfer are shorter for less nuclear evolved donors, which have less massive helium cores.

As described in Sect.~\ref{resultsCVEVOLUTION}, the subsequent evolution can be divided into two different pathways, either with a detached phase or without it.
Irrespective of whether one case or the other happens, the systems will evolve towards shorter orbital periods until they reach the orbital period minimum and start bouncing.
The precise value of the orbital period minimum depends on the donor properties, mainly the mass of its helium core at the onset of mass transfer: the lower the helium core mass, the longer the orbital period minimum.
For donors with extremely-low-mass helium cores, the minimum orbital period can be as long as ${\sim3}$~h.
On the other hand, for those having the highest helium core masses, the orbital period minimum can be as short as ${\sim5}$~minutes.

For \cvs~that evolve through a detached phase, the upper edge and the lower edge and thus the width of the detached phase depend strongly on the helium core mass of the donor at the onset of the first mass transfer phase.
The more massive the helium core of the donor, the longer the detached phase, and in turn the longer (shorter) is the upper (lower) edge of the detached phase. 
In terms of donor entropy and degree of degeneracy, this implies that donors with higher helium core masses at the onset of mass transfer have more time to relax, cool and become more degenerate before mass transfer resumes after the detached phase. 
For this reason, the end of the detached phase corresponds to very short orbital periods in these cases.
On the other hand, donors in \cvs~with continuous mass transfer (i.e. no detached phase) keep their high entropy and remain bloated even during \popm~evolution.
For these systems, the higher the helium core mass of the donor, the lower the entropy (especially near the centre) and for this reason the shorter the orbital period minimum.
The formation of systems like \gaia~and \ztf~can be explained through this channel as should be discussed in more detail in Sects.~\ref{Problem1TR} and \ref{Problem1MB}.

Combining these different tracks implies that at a given orbital period the predicted surface hydrogen abundance is varying substantially during the \popm~evolution.
The shorter the orbital period minimum, the lower the surface hydrogen abundance at a given orbital period.
This is a direct consequence of the chemical profile of the donor at the onset of mass transfer, which is different in all evolutionary tracks shown in Fig.~\ref{FigT}.
Except for AM\,CVn itself, the \cv~channel can explain all the \ams~with reliable parameters as shown in Fig.~\ref{FigT} since the surface hydrogen abundance drops to undetectable levels in the evolutionary tracks passing through them.
One important consequence of the \cv~channel which agrees with the observations is that donors can be substantially larger for their masses in comparison to the white dwarf channel.
This is because of the high entropy of the donor during \popm~evolution due to magnetic braking, which keeps the orbital-angular-momentum-loss timescale much shorter than the thermal timescale of the donor.

To show that the formation channel of \ams~from \cvs~as outlined above does not represent a rare and unlikely evolutionary pathway, we compare the ranges of initial orbital periods of post-\ce~binaries leading to the formation of \hecvs~and \ams.
The initial post-\ce~masses for the tracks shown in Fig.~\ref{FigT} are fixed to $0.8$ (white dwarf) and $1.1$~\Msun~(donor).
Binaries with initial post-\ce~orbital periods from ${\approx2.14}$ to ${\approx2.25}$~d evolve to \hecvs~and remain as \hecvs~at a donor mass as low as $0.01$~\Msun, while those with orbital periods from ${\approx2.25}$ to ${\approx2.64}$~d first evolve to \hecvs~and subsequently evolve to \ams~at a donor mass $\gtrsim0.01$~\Msun.
More details about this transition from \hecvs~to \ams~are provided in Sect.~\ref{Problem1TR}.
The range of orbital periods leading to \ams~is thus broader than that leading to \hecvs.

This result does not depend on the exact value of the initial post-\ce~masses. 
Assuming, for example, post-\ce~binaries hosting initially a white dwarf mass of $0.8$~\Msun~but paired with a companion of mass $1.0$~\Msun, the range of initial post-\ce~orbital periods leading to \hecvs~and subsequently \ams~remains larger than that producing \hecvs~that remain as such at donor masses  ${\geq0.01}$~\Msun.
The resulting orbital period ranges are ${\approx2.52-2.66}$~d and ${\approx2.66-3.00}$~d for \hecvs~and \ams, respectively.  
For systems with initial post-\ce~companions more massive than ${\sim1.2}$~\Msun, for the metallicity (i.e. solar) and treatment of core overshooting we adopted, core overshooting during main-sequence evolution leads to a quick formation of a relatively massive helium core (${\gtrsim0.07}$~\Msun).
Thus, such systems evolve to \ams~irrespective of the white dwarf mass and the orbital period of the initial post-\ce~binary. 
We can thus firmly conclude that it seems more likely to form \ams~with donor masses ${\gtrsim0.01}$~\Msun~from the evolution of \cvs~hosting nuclear evolved donors than \hecvs.

In the above analysis we assumed that the duration of the post-\ce~evolution dominates over the time it takes to form the white dwarf.
This is likely correct because the nuclear time scale of the progenitor of a $0.8$\Msun~white dwarf is much shorter that the post-\ce~evolution we describe here.  
We also implicitly assumed that each initial post-\ce~orbital period below three days is equally likely which represents nothing but a first guess.
More solid relative numbers require to perform binary population synthesis which is beyond the scope of the current paper but will be presented in a future work.

\begin{figure*}
\begin{center}
\includegraphics[width=0.99\linewidth]{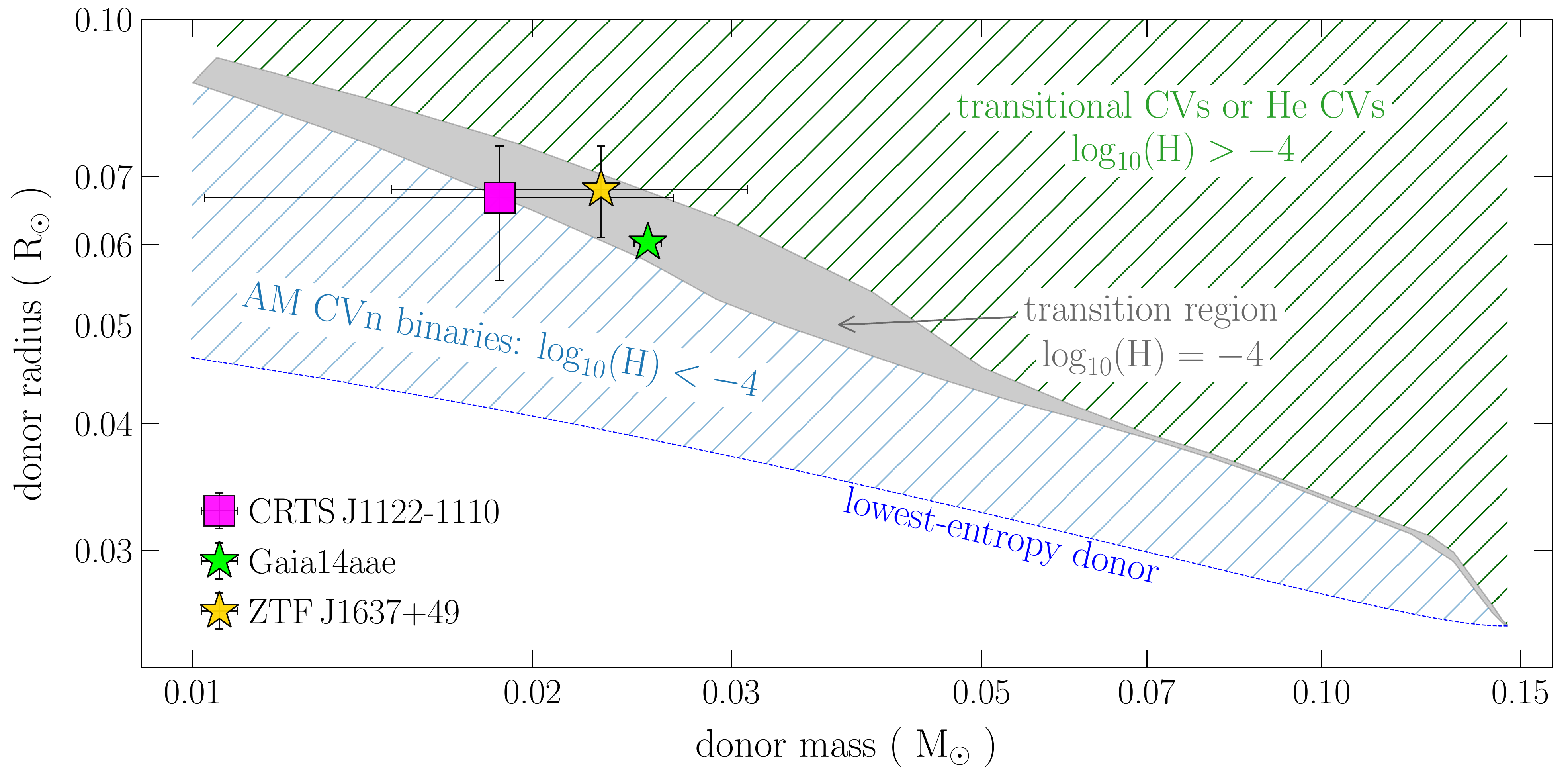}
\end{center}
\caption{
Predicted regions of parameter space for transitional \cvs~and \hecvs~(hatched green region) and \ams~(hatched blue region) as well as the transition region in which \hecvs~evolve into \ams~(grey area).
The transition region is characterized by the evolutionary stages for which the surface hydrogen abundance is $10^{-4}$.
To illustrate this transition region we assumed different initial post-\ce~white dwarf masses ($0.6$, $0.7$, $0.8$, $0.9$ and $1.0$~\Msun) and companion masses ($1.00$, $1.05$, $1.10$, and $1.15$~\Msun), and covered the entire range of orbital periods such that each combination of masses led to the formation of \ams.
Systems located below the transition region have surface hydrogen abundances smaller than $10^{-4}$, irrespective of the initial post-\ce~conditions. In this region our model only predicts \ams~to be found.
This \am~region has a lower limit defined by the radius--mass relation of the lowest-entropy donor in our simulations, which is consistent with the radius--mass relation of a cold, fully degenerate white dwarf.
Systems located above the transition region have surface hydrogen abundances greater than $10^{-4}$ regardless of the initial post-\ce~parameters. According to our model, in this region only \hecvs~and transitional \cvs~should be found.
On the other hand, systems located in the transition region, such as \gaia~(green star), \ztf~(yellow star), and \crts~(magenta square), can have surface hydrogen abundances greater or smaller than $10^{-4}$, depending on the initial post-\ce~properties.
Therefore, our model predicts that in this transition region both \hecvs~and \ams~should exist as it seems to be the case. 
For more details, see Sect.~\ref{Problem1TR}.
}
\label{FigTR}
\end{figure*}

\begin{figure}
\begin{center}
\includegraphics[width=0.99\linewidth]{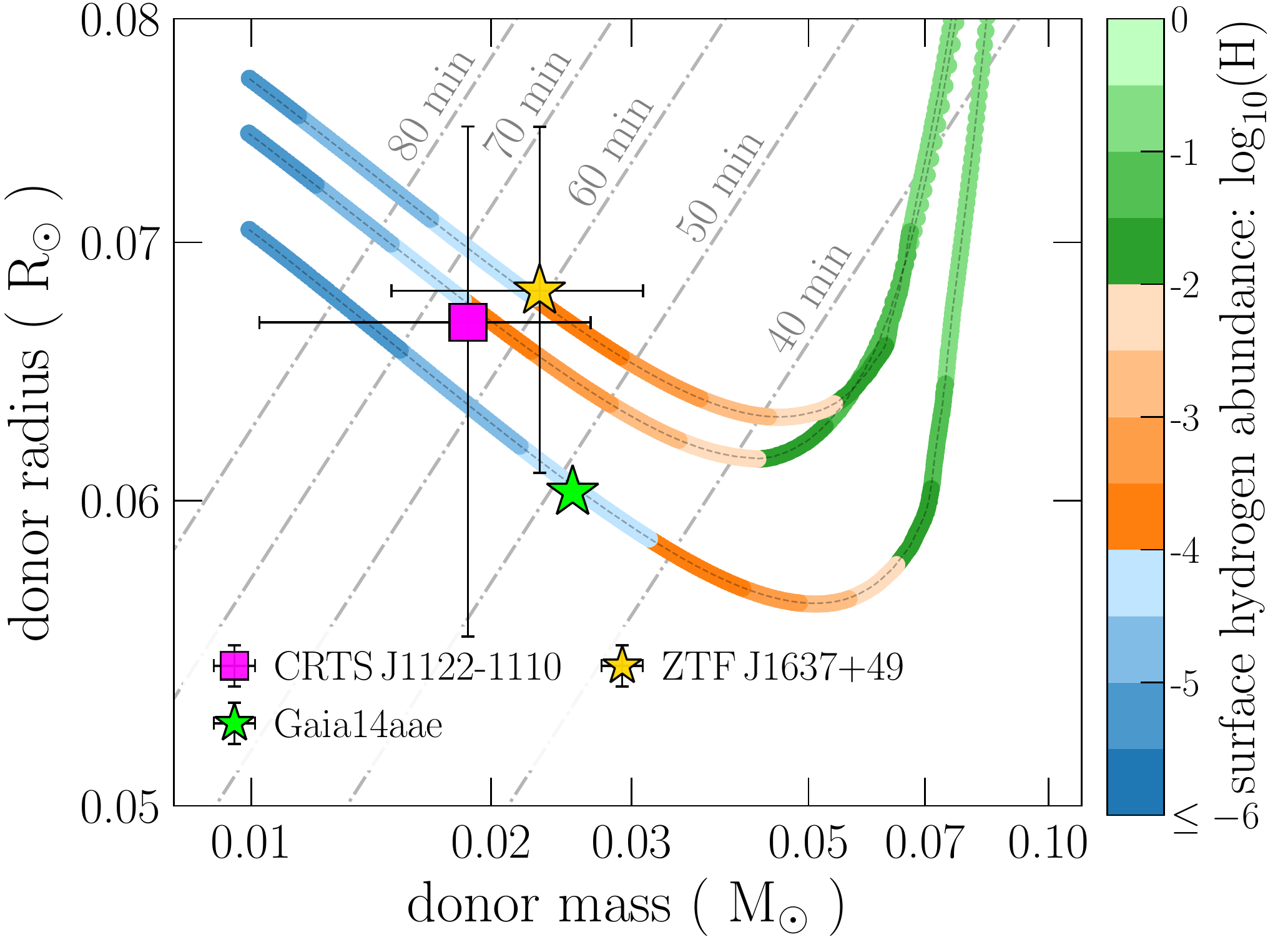}
\end{center}
\caption{
Three evolutionary tracks  around the orbital period minimum  that illustrate the evolution towards \hecvs~and subsequently to \ams~in the plane donor mass versus donor radius.
These evolutionary sequences can explain the observed properties of the \ams~\gaia~(green star) and \ztf~(yellow star) and the \hecv~\crts~(magenta square), which lie within the transition region as shown in Fig.~\ref{FigTR}.
These models have initial post-CE orbital periods, white dwarf and donor star masses, and mass of the helium core of the donor at the onset of mass transfer given by the following values:
$1.14$~d, $0.90$, $1.15$, and ${\approx0.061}$~\Msun~(top track);
$2.495$~d, $1.00$, $1.00$, and ${\approx0.043}$~\Msun~(middle track); and
$1.01$~d, $0.87$, $1.16$, and ${\approx0.070}$~\Msun~(bottom track).
The only difference between the \ams~\gaia~and \ztf~and the \hecv~\crts~is the surface hydrogen abundance, which is still in the detectable regime for the latter.
}
\label{FigTE}
\end{figure}

\begin{figure*}
\begin{center}
\includegraphics[width=0.99\linewidth]{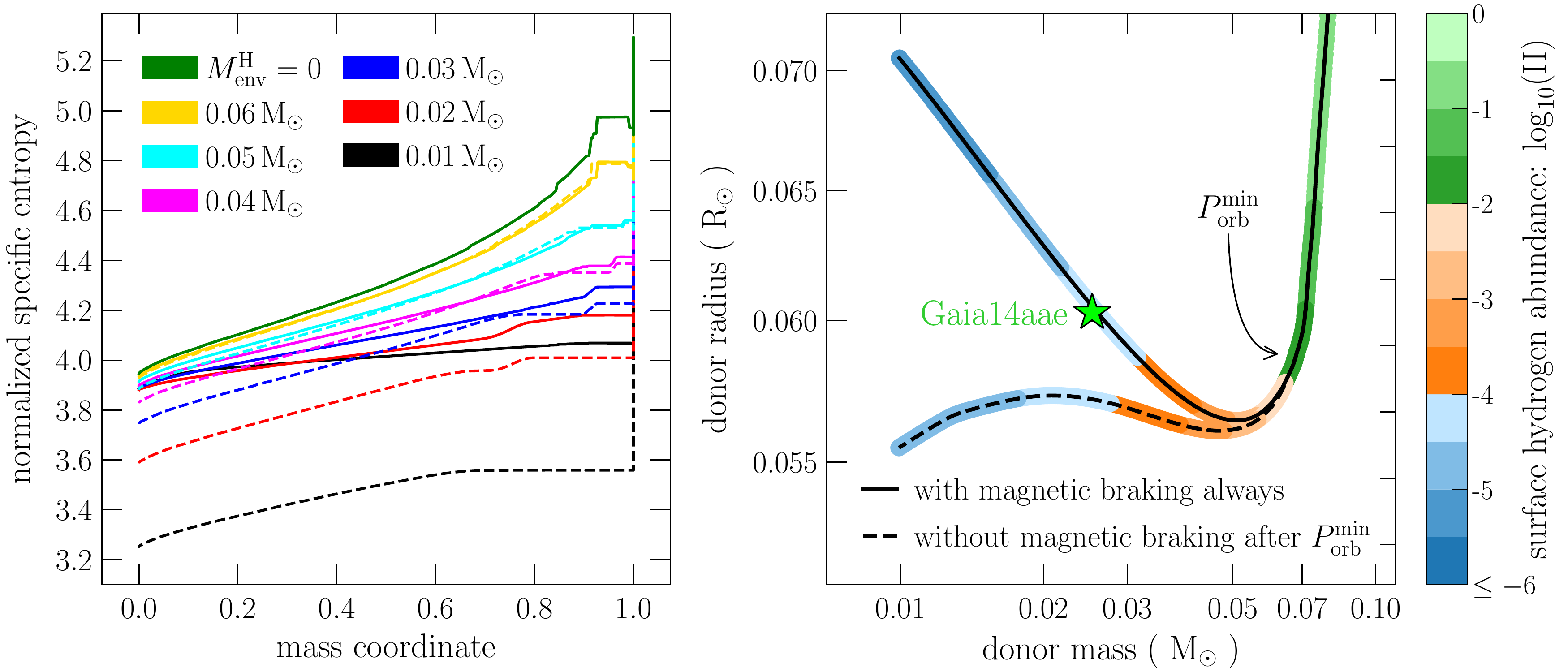}
\end{center}
\caption{
Magnetic braking during \popm~evolution is crucial for explaining systems such as \gaia~(green star), \ztf~and \crts.
We ran two models, one allowing for magnetic braking during \popm~evolution (solid lines) and one assuming no magnetic braking after the hydrogen envelope vanishes at the orbital period minimum (dashed lines).
In both cases, the initial post-\ce~orbital period, white dwarf and companion masses were $1.01$~d, $0.87$ and $1.16$~\Msun, respectively, and the mass of the helium core of the donor at the onset of mass transfer was 0.07~\Msun.
If gravitational wave radiation is the only orbital angular momentum loss mechanism, the existence of \gaia, \ztf, and \crts~cannot be explained because the longer mass-loss timescale causes the entropy of the donor to drop (dashed lines in the left panel) and the donor to relax towards thermal equilibrium.
If, on the other hand, orbital angular momentum loss is sufficiently strong, the central entropy of the donor remains high (solid lines in the left panel) which explains why the donors in \gaia, \ztf, and \crts~are significantly oversized for their masses and orbital periods.
See Sect.~\ref{Problem1MB} for more details.
}
\label{FigO}
\end{figure*}

\subsection{The transition region occupied by \gaia, \ztf~and \crts}
\label{Problem1TR}

Having provided a solution to the first problem by showing that the lack of hydrogen in \ams~can be reproduced by the \cv~channel, we now turn to the discussion on the existence of a region in the parameter space in which both \hecvs~and \ams~co-exist.
To the best of our knowledge, this overlapping region was first identified by \citet{Green_2020} and is occupied by the two \ams~\gaia~and \ztf, and the \hecv~\crts.
All three systems have orbital periods longer than ${\sim50}$~minutes, donors with masses lower than ${\sim0.025}$~\Msun~and radii larger than ${\sim 0.06}$~\Rsun.

The co-existence of \hecvs~and \ams~in this region of the parameter space is nicely explained by the \cv~channel.
As a matter of fact, a region where both types of system co-exist is predicted by this channel and should be actually regarded as a transition region, in which \hecvs~convert into \ams.
In what follows we explain in more detail how \cvs~end up in this region of the parameter space.

In case a \cv~hosts a donor with non-negligible but low helium core mass (${\lesssim0.04}$~\Msun) at the onset of mass transfer, it will become a \hecv~during \popm~evolution and stay as a \hecv~down to donor masses of ${\sim0.01}$~\Msun.
These \hecvs~will likely become \ams~but only at masses smaller than ${\sim0.01}$~\Msun, which are not considered here.
The remaining \cvs~hosting nuclear evolved donors, that is, those with donors having helium core masses ${\gtrsim0.04}$~\Msun~at the onset of mass transfer, first evolve into \hecvs~and stay as such for some time (up to ${\sim50}$~Myr) before becoming \ams.
For these \cvs, depending on the threshold separating \hecvs~from \ams~and on the initial post-\ce~masses, the \popm~system can be recognized as either a \hecv~or an \am.

For a fixed adopted threshold, that is, a unique surface hydrogen abundance separating \hecvs~from \ams~that is adopted for all systems, and a fixed combination of initial post-\ce~white dwarf and companion masses, there is a unique line in the plane donor mass versus donor radius separating \hecvs~from \ams.
Each point on this line corresponds to a donor surface hydrogen abundance equal to the adopted threshold for a given initial post-\ce~orbital period.
The longer the initial post-\ce~orbital period, the larger the donor mass and the smaller the donor radius at which the donor surface hydrogen abundance is equal to the adopted threshold.
This is because the longer the initial post-\ce~orbital period, the more nuclear evolved the donor is at the onset of mass transfer.

Systems above this line in the plane donor mass versus donor radius are predicted to be observed as transitional \cvs~or \hecvs, while those below it should be observed as \ams.
The exact location of the line is only weakly correlated with the initial post-\ce~masses.
In general, the line moves towards higher donor masses, for higher initial post-\ce~donor masses, and towards lower donor masses, for higher initial post-\ce~white dwarf masses.
Combining all the lines resulting from different combinations of initial post-\ce~masses, a region in the plane donor mass versus donor radius arises corresponding to all possible evolutionary stages for which the donor surface hydrogen abundance is equal to the adopted threshold.
This region can be regarded as the transition region, in which \hecvs~are converting to \ams.
An important consequence of the existence of this transition region is that whether a system located inside this region is recognized as a \hecv~or an \am~depends on the initial post-\ce~white dwarf and donor masses.

We show in Fig.~\ref{FigTR} as a grey area the transition region for the threshold we assume, that is, a surface hydrogen abundance of $10^{-4}$, together with the three systems with significantly bloated donors, namely, \gaia, \ztf~and \crts.
We built this region by considering five different initial post-\ce~white dwarf masses ($0.6$, $0.7$, $0.8$, $0.9$, and $1.0$~\Msun) and four different initial post-\ce~donor masses ($1.00$, $1.05$, $1.10$, and $1.15$).
The initial post-\ce~orbital periods were chosen such that, for each combination of initial post-\ce~masses, the entire parameter space for the formation of \ams~was covered.
We have assumed a fixed threshold but as soon as larger samples of \hecv~and \ams~with bloated donors will be available, assuming a range of surface hydrogen abundances might be more realistic.  
The higher the threshold surface hydrogen abundance separating \hecvs~from \ams, the higher the donor mass at which the transition takes place.
For a range of threshold abundances, the transition region would thus be more extended than that shown in Fig.~\ref{FigTR}.

Systems located above the transition region will be observed as either transitional \cvs, or \hecvs, or something in between, since above the transition region the donor surface hydrogen abundance is always ${>10^{-4}}$, irrespective of the initial post-\ce~conditions.
Similarly, systems located below the transition region will be observed as \ams~because the donor surface hydrogen abundance is always ${<10^{-4}}$, regardless of the initial post-\ce~conditions.
The region corresponding to \ams~is limited by the radius--mass relation of a cold, fully degenerate white dwarf, as a donor in an \ams~cannot have a radius smaller than that.
Finally, systems located inside the transition region, such as \gaia, \ztf~and \crts, can be observed as either \hecvs~or \ams.
Whether a particular system is observed as a \hecv~or an \am, that is, whether the surface hydrogen abundance of the donor has dropped below detection levels or not depends on its initial post-\ce~conditions.

The transition region is restricted to donor masses between ${\sim0.01-0.15}$~\Msun.
The lower limit corresponds to the smallest donor masses we investigated.  
It might well be that the transition region extends to lower donor masses but this region is not covered by our simulations.
The upper limit is set by the age of the Universe.
Systems with donor masses greater than ${\sim0.15}$~\Msun~take a very long time to become semi-detached again after the onset of the detached phase so that their total ages become longer than the Hubble time.
Thus, \cvs~with donor masses ${\gtrsim0.15}$~\Msun~when magnetic braking becomes inefficient will be observed as detached double white dwarf binaries.

The extension of the transition region in terms of donor radii depends on the donor masses as shown in Fig.~\ref{FigTR}.
At donor masses ${\gtrsim0.06}$~\Msun, the transition region is very narrow, which is a consequence of the chemical profile of the donor at the onset of mass transfer.
The transition at these masses takes place when the mass of the helium core of the donor at the onset of mass transfer is ${\gtrsim0.08}$~\Msun.
Additionally, the time a given system takes to become an \am~strongly correlates with the donor mass.
The larger the donor mass, the shorter the transition time-scale.
The transition during \popm~evolution takes less than ${\sim10}$~Myr. 
At these relatively large donor masses, the systems above the transition region are still evolving towards the orbital period minimum, so they corresponds to systems that are either transitional \cvs~or \hecvs, while those below are \ams.
Provided the narrowness of the transition region at these high donor masses, we can safely conclude that whether such a system is observed as a \hecv~or an \am~does not depend on the initial post-\ce~conditions.
Instead, the location of the transition for these large donor masses depends exclusively on the evolution of the donor star.

On the other hand, the transition region is sufficiently broad for donor masses ${\lesssim0.06}$~\Msun.
If the mass of the helium core of the donor at the onset of mass transfer is ${\lesssim0.08}$~\Msun, the transition occurs at these donor masses and the transition time-scale can be as long as ${\sim50}$~Myr, for the lowest donor masses.
At these donor masses, whether a system is observed as a \hecv~or an \am~strongly depends on the initial post-\ce~conditions.
Therefore, observed \hecvs~belonging to the transition region, such as \crts, still have enough hydrogen to be detected but are very close to convert to \ams, while observed \ams~inside the region, such as \gaia~and \ztf, have just lost enough hydrogen so that it is currently at a non-detectable level.

Given the importance of the three systems located in the transition region, we show in Fig.~\ref{FigTE} examples of evolutionary sequences that can explain their properties. 
Unfortunately, due to large uncertainties in the measured parameters of \ztf~and the fact that for \crts~the mass ratio is only estimated from the superhump excess--mass ratio relation, discussing the tracks for these two systems in detail represents a rather futile exercise. 
However, as the parameters of \gaia~are sufficiently precise, we will discuss this system and the corresponding evolutionary track in more detail in what follows.

\subsection{The importance of magnetic braking during post-orbital-period-minimum evolution}
\label{Problem1MB}

An important aspect of systems occupying the transition region discussed in Sect.~\ref{Problem1TR} is that they all contain oversized donors.
In fact, the donor radii in these systems are much larger than expected for degenerate objects of the same mass.
Our models naturally explain this.
As an example, we show in Fig.~\ref{FigO} an evolutionary track that can explain the properties of \gaia~assuming an initial post-\ce~orbital period, white dwarf mass and companion mass of $1.01$~d, and $0.87$ and $1.16$~\Msun, respectively, leading to a helium core mass of $0.07$~\Msun, at the onset of mass transfer.
%
%
At the orbital period of \gaia~(i.e. $49.7$~minutes), the white dwarf mass is identical to its initial post-\ce~mass as the mass transfer rate never reaches values high enough to allow for mass growth during \cv~evolution.
The donor mass and radius are $0.0246$~\Msun~and $0.0608$~\Rsun~and the surface hydrogen abundance is $5.19\times10^{-5}$.
These values are in excellent agreement with those derived from observations \citep{Green_2018,Green_2019},
that is, a white dwarf mass of $0.87\pm0.02$~\Msun, donor mass of $0.0250\pm0.0013$~\Msun, donor radius of $0.0603\pm0.0003$~\Rsun, and a surface hydrogen abundance of ${\lesssim10^{-4}}$.
It is important to keep in mind that the shown model is just an example of several tracks that are able to explain \gaia.
Keeping the white dwarf mass fixed to the observed value, we can typically find equally good models for different initial post-\ce~companion masses in the range of ${\sim1-1.16}$~\Msun, as long as the initial post-\ce~orbital period is changed accordingly.

The excellent agreement between observations and theoretical predictions depends crucially on the strong orbital angular momentum loss provided by the CARB model.
According to the latter, \cvs~evolving to the broad portion of the transition region shown in Fig.~\ref{FigTR} never detach, that is, magnetic braking is removing orbital angular momentum throughout the evolution which results in continuous mass transfer, even during \popm~evolution.
This is because a non-negligible portion of the envelope of the donor remains convective and magnetic braking dominates orbital angular momentum loss.
Therefore, the mass-loss timescale remains short enough to keep the donor out of thermal equilibrium, that is, the entropy remains high and the donor substantially bloated.

To illustrate the importance of magnetic braking during \popm~evolution, we ran the same model that provided the excellent fit but this time we arbitrarily suppressed magnetic braking when the hydrogen envelope vanishes at the orbital period minimum.
The corresponding evolution is shown in Fig.~\ref{FigO} by the dashed lines.
Apparently, as soon as magnetic braking turns off, the donor starts to become increasingly degenerate and its radius drops as it loses mass.
At the donor mass of \gaia, the radius is predicted to be significantly smaller than observed, and it drops even further as the evolution continues.

The left panel of Fig.~\ref{FigO} shows entropy profiles of the donors for both simulations and further illustrates the evolution in both cases. 
The first entropy profile corresponds to the moment when the hydrogen envelope vanishes at a donor mass of $0.0644$~\Msun~for both simulations.
At this moment, the donor is only slightly degenerate as its central entropy is high (green lines).
During the subsequent evolution, the entropy profiles significantly change, especially near the centre, and their evolution depends critically on whether orbital angular momentum loss due to magnetic braking is incorporated or not. 
When the donor mass drops to $0.03$~\Msun~(blue lines), both models predict completely different entropy profiles. In case magnetic braking is suppressed, the donor quickly becomes highly degenerate and reaches a very low value near the centre at a donor mass of $0.01$~\Msun~(black dashed line).
This decrease in the central entropy allows the donor to cool and contract and occurs because the mass-loss timescale is long enough. 
In contrast, if magnetic braking is constantly removing orbital angular momentum, the central entropy of the donor remains roughly constant down to a mass of $0.01$~\Msun.
This occurs because the orbital-angular-momentum-loss timescale is always much shorter than the thermal timescale of the donor, which keeps the entropy high and the donor bloated, consistent with the radii inferred from observations.

The strength of orbital angular momentum loss in interacting binaries is frequently estimated based on observational constraints on either the radius of the donor star or the mass transfer rate.   
Both, the donor radius as well as the mass transfer rate depend fundamentally on the orbital angular momentum loss rate.
Orbital angular momentum loss drives mass transfer which in turn drives the donor out of thermal equilibrium. 
We have demonstrated that our model can reasonably well explain the radius of the donor in \gaia, which is caused by efficient magnetic braking during \popm~evolution (Fig.~\ref{FigO}).
The mass transfer rate our model predicts for \gaia~is ${\sim6\times10^{-10}}$~\Msun~yr$^{-1}$.
A robust test for predicted mass transfer rates is provided by the disc instability model.
The critical mass transfer rate separating outbursting and persistent disc systems has been successfully used to constrain other parameters of accreting binaries, even including their distances \citep{schreiber+gaensicke02-1,schreiber+lasota07-1,schreiber2013}.
According to \citet[][their eq.~2]{lasota2008}, the limiting accretion rate for \gaia~(assuming a viscosity parameter of $\alpha=0.1$) is $8.3\times10^{-9}$~\Msun~yr$^{-1}$ which largely exceeds the mass transfer rate predicted by our model which is consistent with the fact that \gaia~is showing outbursts in its light curve \citep{campbell2015}.

While our prediction thus agrees with the solid upper limit that can be derived from the disc instability model, the only estimate for the mass transfer rate in this system derived from observations is an order of magnitude lower than our model predicts, that is, ${\sim3\times10^{-11}}$~\Msun~yr$^{-1}$ \citep{Ramsay_2018}. 
We argue that this mass transfer rate derived from observations most likely underestimates the true mass transfer rate.
First of all, the measured mass transfer rate and the measured donor star radius directly contradict each other. 
It appears impossible that \gaia~could have a mass transfer rate as low as estimated by \citet{Ramsay_2018} and at the same time a donor that is as significantly bloated as measured by \citet{Green_2019}. 
Either the donor radius must be smaller to be consistent with the observationally estimated mass transfer rate, or the mass transfer rate must be higher to be consistent with the observationally estimated donor radius.
Interestingly, if we arbitrarily turn off magnetic braking during \popm~evolution (dashed line in the right panel of Fig.~\ref{FigO}), the donor radius is significantly smaller than the one derived from observations but the predicted mass transfer rate (${\sim2\times10^{-11}}$~\Msun~yr$^{-1}$) 
agrees with the value estimated by \citet{Ramsay_2018}.

As only one of the measurements can be correct, and given that estimates of the mass transfer rates in accreting white dwarf binaries are notoriously uncertain, we strongly believe that in the case of \gaia~the mass transfer rate is higher than has been estimated while the measurement of the donor radius is correct.
The mass transfer rate has been estimated based on modelling the eclipse light curve and subsequent estimates of the contributions from the bright spot.
This approach includes a rather arbitrary assumption of the  spectral response of the bright spot and further assumes that all the overflowing material dissipates a fraction of its kinetic energy in the bright spot, that is, the possibility of stream overflow \citep{schreiber+hessman98} is not considered. 
We believe that combining the uncertainties related to these assumptions can most likely make up 
a difference of one order of magnitude. 
In contrast, deriving the donor star radius from eclipse light curves requires much less assumptions and therefore represents a more reliable measurement.

In any event, even if the donor star radius turns out to be smaller than estimated and the measured mass transfer rate to be correct, we could most likely still explain \gaia.
We would just need a model for which magnetic braking would be much weaker at the orbital period of \gaia.
This would simply require the initial post-\ce~binary to have a longer orbital period so that the donor would have a more massive helium core at the onset of mass transfer.

\begin{figure*}
\begin{center}
\includegraphics[width=0.99\linewidth]{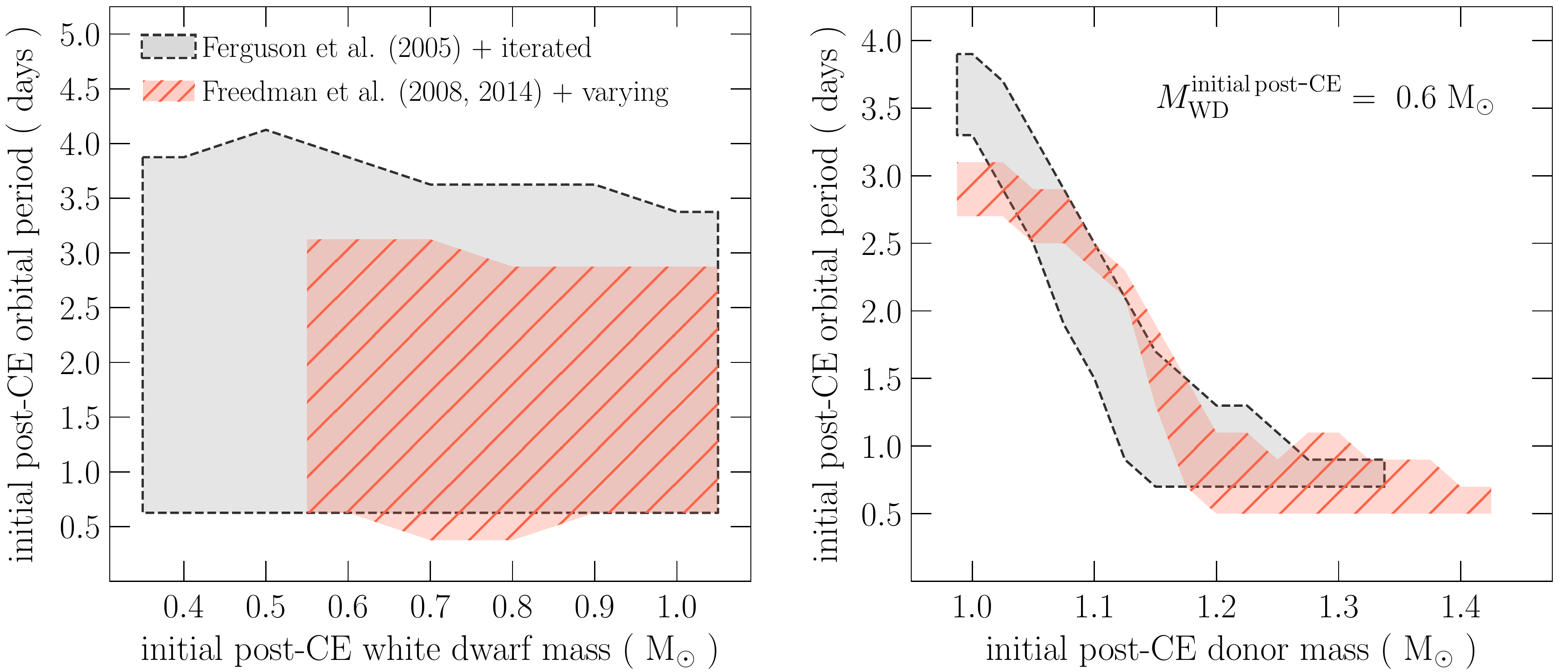}
\end{center}
\caption{
Initial parameters of post-\ce~binaries that first evolve to \cvs~and subsequently to \ams~if the CARB prescription is adopted for magnetic braking for two different choices of opacities and how they are calculated throughout the atmosphere (shaded regions).
To build the four shaded regions, we took into account the uncertainties given by the resolution of our grid of models, corresponding to half the step in white dwarf mass and in orbital period.
Left: Taking into account the full range of considered masses for the companion, systems that evolve into \ams~cover a large range of orbital periods and white dwarf masses.
For our standard assumptions, however, models with initial post-\ce~white dwarf masses lower than ${\sim0.55}$~\Msun~evolve into dynamically unstable mass transfer at the onset of mass transfer and thus do not lead to \ams~(red hatched region). 
If the opacity from \citet{Ferguson2005} coupled with a different treatment of how they are computed is adopted instead, we get dynamically stable mass transfer even for initial post-\ce~white dwarf and companion masses of ${\sim0.4-0.5}$ and $1.3$~\Msun~(grey shaded region).
Thus, if magnetic braking is sufficiently strong, no fine tuning is need to produce \ams~from \cvs.
Right: A finer grid calculated for a fixed white dwarf mass ($0.6$\Msun) illustrates a relation between orbital period and companion mass.
The higher the companion mass, the shorter the orbital period.
See Sect.~\ref{Problem2} for more details.
}
\label{FigF}
\end{figure*}

\subsection{Solving the second problem: fine tuning}
\label{Problem2}

In Sect.~\ref{Problem1} we addressed the problem associated with the hydrogen surface abundances predicted by the \cv~channel and showed that this channel is not only producing \hecvs~as previously thought but also \hecvs~that quickly convert to \ams.
Perhaps somewhat surprising, it seems even more likely to form \hecvs~that subsequently convert to \ams~than \hecvs~that remain as such since the orbital period range from which the latter are formed is narrower than that from which the former are formed.
We now turn our attention to the second frequently mentioned problem of the \cv~formation channel for \ams, that is, the claim that only a small range of initial post-\ce~parameters leads to the formation of \ams~from \cvs~with evolved donors, often called the fine-tuning problem. 
To that end we investigated in more detail the different outcomes of \cv~evolution taking into account a broad range of the initial post-\ce~binary parameter space.

The boundaries separating the different \cv~evolution outcomes are strongly dependent on the combination of masses (white dwarf and companion) as well as on the assumed magnetic braking recipe.
In other words, for a given combination of masses and magnetic braking prescription, there are unique initial post-\ce~orbital periods separating the following outcomes for systems evolving through dynamically stable mass transfer:
%
(i) the \cv~remains as a typical \cv~throughout the evolution;
%
(ii) the \cv~evolves to a \hecv~and remains as such down to a donor mass of ${\sim0.01}$~\Msun;
%
(iii) the \cv~first evolves to a \hecv~and afterwards to an \am~at donor masses ${\gtrsim0.01}$~\Msun;
%
(iv) the \cv~evolves to a detached double white dwarf binary.

A very short initial post-\ce~orbital period leads to outcome (i).
As the initial post-\ce~orbital period increases the \cv~switches its pathway successively from (i) to (iv).
The reason for this is simply that the longer the initial post-\ce~orbital period, the more nuclear evolved the donor, and in turn the more massive the helium core of the donor at the onset of mass transfer.
The transition between the outcomes (iii) and (iv) exists because systems that detach at very long orbital periods may not manage to come into contact again within the Hubble time.

The above listed possible outcomes of post-\ce~evolution require dynamically stable mass transfer. 
In case the donor is initially significantly more massive than the white dwarf or in case the donor is highly evolved, mass transfer can be dynamically unstable.  
Dynamically unstable mass transfer can happen when the donor is initially either a main-sequence star, a sub-giant, or a red giant.
Unless the orbital period at the onset of dynamically unstable mass transfer is very long (i.e. of the order of months or even years), 
the two stars will merge.

To estimate under which conditions a given initial post-\ce~binary evolves into a \hecv~or \am, we calculated a small grid of evolutionary tracks. 
We varied the white dwarf mass from $0.4$ to $1.0$~\Msun, in steps of $0.1$~\Msun, the companion mass from $1.0$ to $1.5$~\Msun, in steps of $0.1$~\Msun, and the orbital period from $0.25$ to $5$~d.
We complemented this coarse but broad grid with a finer gird for the specific 
case corresponding to a white dwarf mass of $0.6$~\Msun.
For the latter grid we used steps of $0.025$~\Msun~for the companion masses and steps of $0.2$~d for the orbital period.
For both grids we defined as \ams~those systems that at some point during their evolution reach a surface hydrogen abundance below $10^{-4}$.
The results for both experiments are shown in Fig.~\ref{FigF} where the shaded regions correspond to the parameter space of initial post-\ce~conditions that leads to the formation of an \am. 
To obtain the range of orbital periods for a given white dwarf mass in the panel on the left (broad but coarse grid), we took into account all companion masses we investigated.

To explore to some extent the impact of stellar evolution assumptions, we investigate two types of low-temperature radiative opacities and how they are computed throughout the atmosphere.
So far in this paper, we adopted the data from \citet{Freedman2008,Freedman2014} and used a varying opacity consistent with the local temperature and pressure throughout the atmosphere, which involves numerical integration of the hydrostatic equilibrium equation.
Models calculated this way produce \ams~for parameters corresponding to the hatched red regions in Fig.~\ref{FigF}.
We also computed models using the data from \citet{Ferguson2005} and a uniform opacity, iterated to be consistent with the final surface temperature and pressure at the base of the atmosphere.
The grey regions in Fig.~\ref{FigF} indicate which initial post-\ce~conditions lead to the formation of \ams~under these assumptions.

The parameter space of initial post-\ce~binaries leading to \ams~is affected by different choices of opacity as well as how they are used to determine the outer boundary conditions.
The parameter space is significantly broader if the \citet{Ferguson2005} opacity is adopted with the iterated procedure in comparison to our standard assumptions.
In particular, for white dwarf masses of $0.4$ and $0.5$~\Msun~in our grid, mass transfer is dynamically unstable for all investigated companion masses if the opacity from \citet{Freedman2008,Freedman2014} is used while the same systems avoid dynamically unstable mass transfer when the opacity from \citet{Ferguson2005} is used (for companion masses as high as ${\sim1.3}$~\Msun).
This is a consequence of the different outer boundary conditions and donor properties (e.g. effective temperature, surface pressure, radius) if this opacity is adopted.
Other parameters are also likely playing a role such as the mixing-length theory, which also affects the star properties and its evolution through the Hertzsprung–Russell diagram.

The main message of Fig.~\ref{FigF}, however, becomes immediately clear. 
If the CARB model for magnetic braking is assumed, there is obviously no fine-tuning problem.
Instead, a broad range of initial post-\ce~binary parameters leads to the formation of \ams. 
This fundamental result appears to be true irrespective of model assumptions for stellar evolution.
Even more convincing, the range of initial post-\ce~orbital periods from which \cvs~give birth to \ams, which are typically between ${\sim0.5}$ and ${\sim4}$~d, are consistent with the observed periods of detached post-\ce~binaries consisting of white dwarfs paired with A--, F--, G-- and K--type main-sequence companions \citep[e.g.][]{Parsons_2015,Hernandez_2021,Hernandez_2022a,Hernandez_2022b}.
Therefore our results indicate that the \cv~channel provides an important pathway to the formation of \ams, which is consistent with recent birth rate estimates \citep{ElBadry_2021}.

The finer grid shown on the right panel of Fig.~\ref{FigF} that has been calculated for a fixed white dwarf mass illustrates that the higher the companion mass, the shorter the orbital period which leads to \ams.
This correlation becomes weaker (maybe even disappears) for companions initially more massive than ${\sim1.2}$~\Msun, which is related to the structure of the core.
We gradually allow core overshooting for companions with masses between $1.1$ and $1.2$~\Msun.
For companions more massive than that, core overshooting is inherent, leading to a significant increase of the core mass which implies that the helium core already starts with a significant mass (${\gtrsim0.08}$~\Msun).
At the same time, these stars have convective cores and radiative envelopes.
Since we apply magnetic braking only when the envelope is convective, the orbital period does not change significantly until the envelope becomes convective, which happens at the onset of helium core formation, when these stars evolve off the main sequence and become subgiants.
For this reason, the initial post-\ce~orbital period has to be short so that these stars fill their Roche lobe when their helium cores are not too massive, which explains
why the correlation between the orbital period and companion mass is somewhat broken when the companion mass is ${\gtrsim1.2}$~\Msun.

\section{Discussion and Prospects}
\label{discussion}

The two most frequently mentioned problems of the \cv~formation channel for \ams~are that it is difficult/impossible to produce systems that do not show any traces of hydrogen and that fine tuning of the initial post-\ce~conditions is required to form \hecvs~or \ams~from \cvs.
We have calculated evolutionary tracks for \cvs~with evolved donor stars assuming magnetic braking according to the CARB prescription \citep{CARB}.
This assumption solves both problems and the formation of \ams~(with undetectable amounts of hydrogen) and \hecvs~is a natural consequence of \cv~evolution.

Our model can easily explain the observed bloated donor stars as well as the more degenerate companions in \ams~and predicts \hecvs~and \ams~to be found in overlapping parts of the parameter space (as observed). 
In what follows we discuss these findings in the context of \am~formation and our understanding of white dwarf binary formation and evolution in general. 
A key ingredient of our model is the assumption that the close binaries continue to experience orbital angular momentum loss due to magnetic braking even after they passed the orbital period minimum.
We start our discussion by taking a more detailed look
at the validity of this assumption before we relate our results to alternative channels for \am~formation and future projects.


\subsection{Magnetic braking in post-orbital-period-minimum systems}

We have shown in Sect.~\ref{Problem1MB} that magnetic braking as additional source of orbital angular momentum loss during \popm~evolution is crucial to reproduce systems such as \crts, \gaia~and \ztf.
A key result of our simulations is that the entropy of the donors in 
these systems is kept high which means that they are only slightly degenerate (left panel of Fig.~\ref{FigO}) and we therefore think that it is reasonable to assume magnetic braking to be active for these systems.

The main conditions for the existence of efficient orbital angular momentum loss due to magnetic braking can be summarised as follows: (i) the donor must be rapidly rotating, (ii) the donor must have a non-negligible magnetic field, which requires a non-negligible convective envelope and (iii) the donor must lose mass in a wind, which also must leave the binary.
We believe these three conditions are satisfied. 
In case of \cvs~evolving to \ams, the donor spin is always synchronised with the orbital motion, even during \popm~evolution, which guarantees that condition (i) is met.
If the helium core mass of the donor is lower than ${\sim0.1}$~\Msun~at the onset of mass transfer, condition (ii) is met because the donor has a non-negligible convective envelope 
throughout the evolution and consequently a magnetic dynamo should generate the required magnetic field.
Winds of low-mass stars are in general driven by the magnetic field itself, unlike the winds of hot luminous stars which are radiatively driven.
The escape velocity for the bloated donors in systems such as \crts, \gaia~and \ztf~does not exceed that of low-mass stars. 
According to Eq.~\ref{Eq:MB-CARB}, the required mass-loss rate in the form of fast winds is moderate, of the order of ${10^{-17}-10^{-15}}$~\Msun~yr$^{-1}$.
It appears therefore reasonable to assume that the magnetic fields we expect to be generated in \popm~systems are equally able to drive wind mass loss.

Interestingly, understanding standard \cv~evolution also seems to require extra orbital angular momentum loss, not only below the orbital period gap, but also during \popm~evolution.
\citet{Knigge_2011_OK} showed that to reproduce the observed orbital period minimum \citep[${\approx79}$~minutes,][]{McAllister_2019} an extra source of orbital angular momentum loss corresponding to ${\sim1.5}$ times that of gravitational wave radiation is required.
Similarly, the estimated properties of \popm~systems can only be explained if the orbital angular momentum loss rate is higher than that provided by emission of gravitational waves \citep{palaetal_2017,Pala_2020,Pala_2022}.
Magnetic braking could be the physical mechanism behind the required extra orbital angular momentum loss during \popm~evolution.

The only way we can think of to further constrain magnetic braking in \cvs~with evolved donors would be to establish a representative (ideally complete volume limited) and well characterised sample of these systems.
This would allow to confront model predictions with, for instance, the observed orbital period distribution and/or mass transfer rates (although this is difficult to measure) as a function of orbital period. 
This sort of comparison is powerful and can lead to useful constraints as has been shown previously for detached post-\ce~binaries and \cvs~\citep[e.g.][]{Gansicke_2009,schreiberetal10-1,Knigge_2011_OK,Schreiber_2016, Belloni_2020a,Pala_2022}.


\subsection{Comparison with previous works}
\label{Sarkar}

Most previous studies dedicated to the \cv~channel for the formation of \ams~\citep[e.g.][]{Podsiadlowski_2003,Nelemans_2010,Goliasch_2015,Kalomeni_2016,Liu_2021} assumed the relatively weak magnetic braking prescription proposed by \citet{RVJ}.
Despite assuming the same angular momentum prescription, their main conclusions concerning the importance of the \cv~channel were substantially different. 
\citet{Podsiadlowski_2003} carried out a thorough binary population synthesis study and found that the \cv~channel provides a potentially important channel for the formation of \ams, being competitive with the other channels.
Similarly, \citet{Liu_2021} investigated the formation of \ams~that are LISA sources and found that the \cv~channel plays an important role in forming \am--LISA sources in the Milky Way.
On the other hand, \citet{Goliasch_2015} and \citet{Kalomeni_2016} found that the formation of \ams~through this channel requires a finely tuned set of initial post-\ce~conditions, which makes the contribution of this channel to the intrinsic population of \ams~negligible.

While some of the above listed investigations concluded that the role played by the \cv~channel in the formation of \ams~cannot simply be ignored, in most of these early works the surface abundances was not considered.
Thus, it is not clear from these studies which fraction of the predicted \am-like population are indeed \ams, that is, systems lacking detectable amounts of hydrogen, and which ones are actually \hecvs.
The first who investigated the chemical composition of donors in \ams~expected from each channel were \citet{Nelemans_2010}.
They concluded that the donors predicted by the \cv~channel typically have hydrogen abundances that should be detectable.
This implies that, assuming relatively weak magnetic braking, the most likely outcome of \cv~evolution with nuclear evolved donors are \hecvs, not \ams.

We found that sufficiently strong magnetic braking, such as that provided by the CARB prescription, can overcome any potential fine-tuning problem, as illustrated in Fig.~\ref{FigF}.
This happens because high orbital angular momentum loss rates allow \cvs~hosting more nuclear evolved donors to evolve towards shorter orbital periods.
This in turn expands the initial post-\ce~parameter space from which \ams~form.
Assuming strong magnetic braking also guarantees that the surface hydrogen abundance of donors during \popm~evolution drops to undetectable levels, as illustrated in Figs.~\ref{FigX}, \ref{FigT}, \ref{FigTE} and \ref{FigO}.
We can therefore conclude that the main problems faced by the \cv~channel were model-dependent as they arose from the assumption of weak magnetic braking.
Most importantly, our results clearly indicate that \ams~formed through the \cv~channel may be potentially numerous, and the \cv~channel may be the dominant formation channel of these systems, unlike previous expectations.

The \cv~channel under the assumption of strong magnetic braking was recently also investigated by \citet{Sarkar_2023a}, from now on 
\citetalias{Sarkar_2023a}, who adopted their own magnetic braking prescription \citep{Sarkar_2022}. 
In addition to assuming a different prescription for strong magnetic braking, the work of 
\citetalias{Sarkar_2023a} is fundamentally different to ours in several other important aspects.

First, we calculated evolutionary tracks down to the masses and radii of the observed systems while the code of \citetalias{Sarkar_2023a} does not allow to remove mass from degenerate regions and is therefore  unable to follow the evolution when the donor mass drops below ${\sim0.03}$~\Msun. 
For comparison with observed systems \citetalias{Sarkar_2023a} therefore extrapolated their evolutionary tracks 
which might not be correct. The response of the donor star to mass transfer may change because of changes in its structure or (potentially related) changes in the orbital angular momentum loss rate as illustrated in our Fig.~\ref{FigO} which might not be easy to extrapolate\footnote{Taking a closer look, it seems that the extrapolation of the red track in fig.~10 of \citetalias[][their model~1,]{Sarkar_2023a} does not follow the slope of the track. Instead there seems to be a kink when the extrapolation connects to the calculated track at ${\sim0.03}$~\Msun.}.

Second, in contrast to our simulations, \citetalias{Sarkar_2023a} could not obtain the threshold separating convergent from divergent evolution.
As explained in their work, this is a shortcoming of their simulations and can be attributed to the increasing degree of degeneracy of the core of the donor when it evolves from the subgiant phase to the red-giant phase.
It is therefore not clear under which conditions their evolutionary tracks are indeed convergent and produce \ams~which, in our opinion, makes it difficult (if not impossible) to reliably estimate the relative number of initial post-\ce~binaries that evolve into \ams~(and thus the importance of the \cv~channel in general) based on their simulations.

Unsurprisingly, given the very different approaches and levels of detail, also the results obtained by 
\citetalias{Sarkar_2023a} and by us are very different. 
The evolutionary tracks calculated by 
\citetalias{Sarkar_2023a} could not explain the long-period \ams~with significantly bloated donors such as \gaia~and \ztf, in sharp contrast to our findings.
The same holds for systems with low-entropy donors such as GP~Com and ZTF~J0407--00, which are naturally explained by our tracks but cannot be reproduced by \citetalias{Sarkar_2023a}.
As \citetalias{Sarkar_2023a} do not provide enough details about their models of the donor star, it is impossible for us to understand the reason behind the different predictions. 
We assume that they are related either to different assumptions concerning opacities, outer boundary conditions, and mixing theory, or caused by the fact that \citetalias{Sarkar_2023a} extrapolate their evolutionary tracks (including the hydrogen abundances).

Another striking difference is that the magnetic braking prescription assumed by \citetalias{Sarkar_2023a} does not lead to a detached phase when the donors are more nuclear evolved at the onset of mass transfer.
For this reason, it seems unlikely that the `terminal' \cvs~identified by \citet{ElBadry_2021} can be explained by their model.
Recent results indicate that the detached phase might be fundamental for explaining the existence of some detached double white dwarfs \citep{Chen_2022} and millisecond pulsars with extremely-low-mass white dwarf companions \citep{Istrate_2014,Chen_2021,Soethe_2021}.   
More generally, it is not clear if the model of \citetalias{Sarkar_2023a} can explain the observed population of CVs hosting nuclear evolved donors, transitional \cvs, or \hecvs~since these authors compared their evolutionary tracks only with \cvs~hosting unevolved donors (which are clearly not the progenitors of \ams) and \ams.

Of course, also our general conclusion differs from that of \citetalias{Sarkar_2023a}. 
While they concluded that the \cv~channel is still less important than the white dwarf and helium star channels, we suggest the opposite. We believe that our result is more reliable because our simulations provide information on which tracks converge and which ones diverge and calculate the structure of low-mass donor stars. 
However, we admit that also our work does not provide final evidence for the dominance of the \cv~channel.
This would require proper binary population synthesis which is beyond the scope of this paper but will be presented in an upcoming article. 
Finally, following that long list of differences and disagreements, we close this section by stating 
that the evolutionary tracks calculated by \citetalias{Sarkar_2023a} were successful in explaining the existence of systems such as YZ~LMi, V396~Hya, CR~Boo and HP~Lib, which is in perfect agreement with our results.

\subsection{Alternative formation channels}

We have shown that the formation of \ams~and \hecvs~from \cvs~with evolved donors is a natural consequence of binary evolution if the CARB prescription for magnetic braking is assumed.
This model can explain the characteristics of all \ams~with reliable parameters, except for AM\,CVn itself.
In addition, no fine tuning of the initial post-\ce~binary parameters is required to produce \ams~from \cvs. 
Provided that it cannot explain AM\,CVn itself, this of course does neither imply that our model is the only model that can explain the formation of \ams~nor that it is the one most frequently chosen by nature. 
Additionally, if the radii and masses of \ams~derived from the superhump excess--mass ratio relation are correct, our model is unable to explain some of the systems with donors that are sufficiently massive and large, such as SDSS J1908$+$3940 \citep[][and references therein]{BK2021}.

The existence of these systems might imply that alternative formation channels are at work. 
Alternative channels proposed for \am~formation are the white dwarf and the helium star channel.
Population synthesis have been performed for both channels \citep[e.g.][]{Nelemans_2001b,Liu_2022} and the predicted birth rates roughly agrees with those derived from observations, although this conclusion depends on unconstrained parameters in the theoretical modelling (such as the \ce~efficiency) and assumptions on observational selection effects and/or biases. 
A strong argument in favour of the \cv~channel is the fact that these alternative channels are unable to predict the formation of \hecvs.

There is overwhelming evidence that the white dwarf channel can neither produce \hecvs~nor is it easy to form \ams~with bloated donor stars in significant numbers because the last episode of mass transfer, which must be \ce~evolution, would need to produce a binary with an orbital period such that the donor is virtually filling its Roche lobe immediately after \ce~evolution.
Otherwise, the white dwarf entropy would quickly drop after the white dwarf formed \citep[e.g.][]{Wong_2021}.
Even in such cases, present-day systems hosting bloated donors such as \gaia~and \ztf~could not be explained by the white dwarf channel \citep[][their fig.~12]{vanRoestel_2022}.
This is simply because in order to explain such systems, the newly born white dwarf should have a mass lower than $0.1$~\Msun, which is not possible as an outcome of \ce~evolution unless a thus far unknown source of energy helps to expel the \ce.

\citet{Sarkar_2023b} intended to circumvent these problems by assuming (i) an extremely efficient \ce~evolution for donors on the subgiant or early giant branch, (ii) that the pre-white dwarf is filling the Roche-lobe directly after \ce~evolution and (iii) that strong magnetic braking is keeping the entropy of the donor high\footnote{The term `helium star channel' traditionally refers to systems containing (at least for some time) a naked helium burning star donor \citep[e.g.][]{Nelemans_2010,Solheim_2010}. In the channel proposed by \citet{Sarkar_2023b}, the donor mass is too small for helium burning to ignite. Therefore, their donors should be called semi-degenerate pre-helium white dwarfs \citep[e.g.][]{maxtedetal14-1} and the corresponding evolution should be considered as part of the white dwarf channel.}.
Assumption (iii) is a good idea which we independently developed and incorporated in the simulations of the \cv~channel we present in this paper.
Assumption (ii) certainly requires some fine-tuning but in principle some (most likely very few) systems could evolve exactly in that way if (and only if) assumption (i) was reasonable.
However, \ce~evolution as efficient as assumed by \citet{Sarkar_2023b} appears to be extremely unlikely if recent observational and theoretical works are taken into account.

A highly efficient \ce~phase for subgiants or slightly evolved giants would predict the existence of close detached binaries consisting of extremely-low-mass white dwarfs with M-dwarf companions.
Such systems would be easy to identify in spectroscopic surveys (such as the Sloan Digital Sky Survey) because the relatively large white dwarfs would dominate the blue part of the spectrum.
The absence of extremely-low-mass white dwarfs in observed samples of post-\ce~binaries \citep[][their fig.~1]{rebassa-mansergasetal11-1} therefore clearly argues against very efficient \ce~evolution.
In fact, the very same absence of extremely-low-mass white dwarfs among white dwarf plus M dwarf binaries is one of the reasons why many 
groups independently find small values of the \ce~efficiency (about one third) to be more likely \citep[e.g.][]{Zorotovic_2010,toonen+nelemans13-1,Camacho_2014,Cojocaru_2017,Scherbak_2023}.

Interestingly, extremely-low-mass white dwarfs and pre-white dwarfs are found in detached systems with more massive secondary stars and have been shown to be consistent with resulting from dynamically stable mass transfer, based on both theoretical \citep[e.g.][]{chenetal17-1,lietal19-1} as well as observational results \citep{lagosetal20-1,parsonsetal23-1}.
A combination of \ce~evolution (with a small efficiency) and dynamically stable mass transfer therefore produces populations in good agreement with the observations.   
It appears to be impossible to obtain a similar agreement assuming the extremely efficient \ce~scenario assumed for subgiant or early giant donors proposed by \citet{Sarkar_2023b} as this scenario would predict the existence of extremely-low-mass (pre-)white dwarfs independent of the secondary mass.
In agreement with \citet{Wong_2021} and \citet{vanRoestel_2022} we can therefore exclude the white dwarf channel to produce long-period \ams~or \hecvs~with bloated donors.

An additional problem of the white dwarf channel has been identified by \citet{Shen_2015} who discovered that the initial phase of hydrogen-rich mass transfer leads to a nova-like eruptions on the accreting white dwarf.
Dynamical friction within the expanding nova shell may then lead to consequential angular momentum loss which likely causes the binary to merge.
This result does not change even if only ten per cent of the energy required to unbind the nova shell come from orbital energy.
This makes the double white dwarf channel further unlikely to play a major role in \am~formation.

It is important to note, however, that if indeed a non-negligible fraction of orbital energy is required to expel the nova shell after the onset of mass transfer in the double white dwarf channel, the very same effect would lead to unstable mass transfer when the detached white dwarf binaries predicted by our model (the detached post-\cvs~and pre-\ams) start mass transfer again. 
While this would reduce the number of \ams~predicted by our channel, it would not affect the ability of our model to explain observed \ams~as systems that just avoid the detached phase cover virtually the same parameter space (see Fig.~\ref{FigT}).

For the second alternative channel, the helium star channel, we cannot exclude in principle that it is possible to form long-period \ams~with significantly bloated donor stars. 
The only work dedicated to \ams~that addressed this channel and followed the evolution to donor masses as low as $0.01$~\Msun~is that by \citet{Wong_2021}.
These authors investigated only one system and showed that the donor entropy also drops and the donor cools down and contracts when the mass is sufficiently low, at orbital periods ${\gtrsim40}$~minutes (their fig.~1).
We conclude that, for the time being, the \cv~channel is the only formation channel that predicts the existence of long-period \ams~and \hecvs~with bloated donors with similar donor masses, and therefore most likely plays a major role in the formation of \ams~and \hecvs.


\subsection{Future work}

Our simulations show that the \cv~formation channel can not only explain the existence of \hecvs~but also that of \ams~and \ams~with bloated donor stars.
In addition, the range of initial post-\ce~binary parameters that lead to the formation of these systems is rather broad, which means that no fine-tuning of the initial post-\ce~conditions is required.   
However, we are well aware of the fact that our simulations depend to some degree on assumptions concerning still uncertain parameters and processes during stellar evolution.
Testing dependencies of model predictions on assumptions concerning the outer boundary condition, opacity, mixing, chemical diffusion, or rotation represents a natural next step towards a better understanding of the \cv~formation channel.
While representing a necessary exercise, most likely, the parameter space of initial conditions for post-\ce~binaries that evolve into \hecvs~and~\ams~may slightly depend on the above-mentioned assumptions, but we expect the model to robustly predict \ams~and \hecvs~independently of these details of stellar structure and evolution.

A probably much more constraining test for the \cv~formation channel, would consist in performing dedicated binary population synthesis and comparing model predictions concerning the entire population of \ams~and \hecvs~with a representative sample of observed systems. 
Performing theoretical binary population models using our model for \cv~evolution is clearly possible and we intend to do so in the near future. 
The most convenient approach will be to use rapid binary evolution codes such as \bse~\citep{Hurley_2000,Hurley_2002} to simulate the evolution of main sequence binary stars and \ce~evolution.
Using such a fast code to describe the formation of post-\ce~binaries will allow us to study the impact of model assumptions that are critical such as the \ce~efficiency, stability of mass transfer, and the zero-age binary population.
For the subsequent evolution, we will have to use the MESA code adopting the scheme we presented in this work to produce correct model predictions.

Interestingly, performing these binary population models will allow us to compare our predictions with observed samples of four different white dwarf binary stars: detached post-\ce~binaries with intermediate mass companions, \cvs~with evolved donors, \ams~and \hecvs, and detached double white dwarfs containing extremely-low-mass white dwarfs (because of the extended detached phase we predict for binaries with sufficiently evolved donors at the onset of mass transfer).
The latter are important sources of low-frequency gravitational waves \citep{Chen_2022}.
While so far for none of these white dwarf binaries a complete volume limited sample has been established, thanks to Gaia and photometric variability surveys \citep[e.g.][]{vanRoestel_2022,ElBadry_2021}, such samples might be within reach in the near future.

\section{Conclusions}
\label{conclusions}

We have carried out dedicated binary evolution models of the \cv~channel for the formation of \ams~using the \mesa~code.
The main criticisms this evolutionary channel was facing in the past were that it is unable to produce \ams~with undetectable amounts of hydrogen (a defining feature of \ams) and that only a small part of the initial post-\ce~binary parameter space leads to the formation of \ams~or \hecvs.
In other words, to make the channel work fine tuning of the initial post-\ce~conditions is required for \cvs~to evolve into \ams.
Both these arguments against the \cv~formation channel are model-dependent. 
Adopting the CARB model for magnetic braking, which provides sufficiently strong orbital angular momentum loss, \cv~evolution with more nuclear evolved donors is convergent, leading to more hydrogen-deficient \am~donors.
As more nuclear evolved donors are bigger and thus necessarily require longer orbital periods prior to the onset of mass transfer, also the range of initial post-\ce~conditions that lead to the formation of \ams~significantly increases which solves the fine-tuning problem. 
Our model can explain the characteristics of virtually all observed \ams~with reliable parameters including those with bloated donors. The only exception so far is AM\,CVn itself which has likely formed through one of the alternative channels. 
While our results indicate that \cvs~most likely play an important role in the formation of \ams, further calculations are still needed to address how stellar evolution assumptions affect the evolutionary tracks.
We also plan to perform proper binary population synthesis for \am~formation from \cvs~assuming the CARB model for magnetic braking to estimate the relative importance of the \cv~formation channel for the observed \am~population.


\begin{acknowledgements}

We thank the Kavli Institute for Theoretical Physics (KITP) for hosting the program ``White Dwarfs as Probes of the Evolution of Planets, Stars, the Milky Way and the Expanding Universe''.
This research was supported in part by the National Science Foundation under Grant No. NSF PHY-1748958.
We thank Matthew J.~Green and S.~Wong for pleasant and helpful discussions during the KITP program.
DB acknowledges financial support from {FONDECYT} grant number {3220167}.
MRS was supported from {FONDECYT} grant number {1221059} and ANID, – Millennium Science Initiative Program – NCN19\_171.

\end{acknowledgements}

%
%

\bibliographystyle{aa} 
\bibliography{references} 

\end{document}